\documentclass[aps,amsmath,amssymb,twocolumn,showpacs,prb]{revtex4-1}
\usepackage{graphicx,color}
\graphicspath{{./}{./figures/}}
\usepackage{dcolumn}
\usepackage{bm}

\definecolor{labelkey}{cmyk}{.4,.2,0,0}

\newcommand{\half}{\frac12}
\newcommand{\red}{\color{black}}

\newcommand{\BEQ}{\begin{equation}}
\newcommand{\EEQ}{\end{equation}}
\newcommand{\bea}{\begin{eqnarray}}
\newcommand{\eea}{\end{eqnarray}}
\newcommand{\be}{\begin{equation}}
\newcommand{\ee}{\end{equation}}
\newcommand{\BEA}{\begin{eqnarray}}
\newcommand{\EEA}{\end{eqnarray}}

\newcommand{\ra}{\right\rangle}
\newcommand{\la}{\left\langle }
\renewcommand{\d}{{\rm d }}

\newcommand{\p}{\partial}
\newcommand{\nn}{\nonumber }

\newcommand{\rme}{\mathrm{e}}
\newcommand{\rmd}{\mathrm{d}} 

\begin{document}

\title{Equilibrium avalanches in spin glasses}
\author{Pierre Le Doussal$^*$, Markus M\"{u}ller$^\dagger$, and Kay J\"org Wiese$^*$} 
\affiliation{$^*$CNRS-Laboratoire
de Physique Th{\'e}orique de l'Ecole Normale Sup{\'e}rieure, 24 rue
Lhomond,75005, Paris, France} 
\affiliation{$^\dagger$The Abdus Salam International Centre for Theoretical Physics, P. O. Box 586, 34151 Trieste, Italy}
\date{\today}

\begin{abstract}
We study the distribution of equilibrium avalanches (shocks) in Ising spin glasses which occur
at zero temperature upon small changes in the magnetic field. For the infinite-range Sherrington-Kirkpatrick model we present a
detailed derivation of the density $\rho(\Delta M)$ of the magnetization jumps $\Delta M$. It is obtained
by introducing a multi-component generalization of the Parisi-Duplantier equation, which
allows us to compute all cumulants of the magnetization.  We find that $\rho(\Delta M) \sim \Delta M^{-\tau}$ 
with an avalanche exponent $\tau=1$ for the SK model, originating from the marginal stability (criticality) of the
model. It holds for jumps of size $1 \ll \Delta M < N^{1/2}$ being provoked by changes of the external field by $\delta H = O(N^{-1/2})$ where $N$ is the
total number of spins. Our general formula also suggests {\red that} the  density {\red of} overlap
 $q$ between initial and final state in an avalanche is $\rho(q) \sim 1/(1-q)$. These results show interesting
similarities with numerical simulations for the out-of-equilibrium dynamics of the SK model.
For finite-range models, using droplet arguments, we obtain the prediction $\tau= (d_{\mathrm{f}} + \theta)/d_{\mathrm{m}}$ 
where $d_{\mathrm{f}},d_{\mathrm{m}}$ and $\theta$ are the fractal dimension, magnetization exponent and
energy exponent of a droplet, respectively. This formula is expected to apply to other glassy disordered systems, such as the random-field model and
pinned interfaces. We make suggestions for further numerical investigations, as well as experimental studies of the Barkhausen noise in spin glasses. 

\end{abstract}
\maketitle


\section{Introduction}
The low-temperature response of disordered systems often proceeds in jumps and avalanches.~\cite{crackling,granular,earthquakes,fracture,LeDoussalWieseMoulinetRolley2009,vortex,mwalls,barkhausen,eglass}
These processes are beyond standard thermodynamic calculations and are therefore relatively difficult to access and describe analytically
\cite{soc,AvalanchesRFIM,avalanchesFRG,LeDoussalMiddletonWiese2008,RossoLeDoussalWiese2009a,vives}. 
In a recent article~\cite{us}, we succeeded in calculating the statistics of equilibrium avalanches (also called shocks) in a variety of disordered systems described by mean-field theory based on Parisi replica symmetry breaking. This encompasses in particular the canonical Sherrington Kirkpatrick (SK) model for the Ising spin glass \cite{SK,MPV} and elastic manifolds in the limit of a large number of transverse dimensions. Although it has been known for a while that the equilibrium magnetization curve $M(H)$ of the SK model undergoes a
sequence of small jumps as $H$ is increased \cite{NonAvOfsuscept}, their statistics had not been obtained previously. The aim of the present article is to provide a detailed derivation of the distribution of avalanche sizes for the SK model. We introduce replica techniques that significantly extend the formalism developped in Ref.~\onlinecite{BMP} to study velocity correlations in high-dimensional Burger's turbulence. It also
generalizes previous studies of the variance of equilibrium jumps to their full distributions \cite{FRGRSB,FRGlargeN,RizzoYoshino}.
We expect this technique to be useful in several other contexts as well. In particular, it should be helpful to describe the response of complex systems to a small change of parameters, a problem that arises in a variety of fields ranging from condensed-matter physics of complex systems, optimization problems to econophysics \cite{vertexcover,coloring,1stepstabilitykSAT,jpb}.

The main result of our calculation is that the distribution of jumps takes a scale-free form, described by a power law of the jump size. This is intimately tied to the criticality of the spin-glass phase of the models analyzed \cite{criticalityMF}, and we conjecture that such a criticality is a feature of a large variety of frustrated glassy systems.~\cite{Palassini} The exact result obtained in the SK model finds a natural interpretation which allows for an extension to finite dimensions via droplet scaling arguments. Those relate the equilibrium-avalanche exponent to critical properties of droplet excitations.

Our results complement previous numerical simulations by Pazmandi et al.~\cite{SKnumerics} 
on out-of-equilibrium hysteresis at $T=0$ in the SK model, which exhibit surprising similarities, as we will discuss. Understanding the relations between these results requires further numerical investigations of dynamic and static avalanches, both in mean-field and finite-dimensional spin glasses.
Our results suggest to look for power-law distributed Barkhausen-type noise in spin and electron glass experiments, as will be discussed.

This paper is organized as follows:
In Sec.~II we revisit the Parisi saddle-point equations in the presence of a small varying external magnetic field. In Sec. III, we generalize the Paris-Duplantier equations to compute the moments of the magnetizations in different fields. From that calculation we extract the distribution of equilibrium jumps  in Sec.~IV. In Sec.~V we consider the case of finite-dimensional spin glasses, and using droplet arguments we obtain a power-law distribution of equilibrium avalanches.
In Sec.~VI we discuss the connection with previous numerical studies on spin and electron glasses, and propose experimental  and numerical investigations.

\section{Model and method}

\subsection{Model and aim of the calculation}

We study the SK spin-glass model of energy
\begin{equation}
\label{SK}
{\cal H} = - \sum_{i,j=1}^N J_{ij} \sigma^i \sigma^j - H_{\mathrm{ext}} \sum_{i=1}^N \sigma^i,
\end{equation}
where the $J_{ij}$ are i.i.d.\ centered Gaussian random variables of variance $J^2/N$, that couple all $N$ Ising spins, and $H_{\mathrm{ext}}$ is the external field.

Our aim is to follow the equilibrium state as a function of the applied field $H_{\mathrm{ext}}$ at low temperature $\beta^{-1}=k_{\mathrm{B}}T =T\ll J$. We consider small variations of the applied field around a reference value $H$, $H_{\mathrm{ext}} = H + \frac{h}{\sqrt{N}}$. 

We are interested in the total magnetization in a given sample,
\begin{equation}
M(H_{\mathrm{ext}}) = \sum_i \left<\sigma^i\right>_{H_{\mathrm{ext}}} = - \partial_{H_{\mathrm{ext}}} F\ ,
\end{equation}
where {\red $F=-k_{\mathrm{B}}T\ln{\rm Tr} \exp(-\beta  {\cal H})$} is the free energy. Since upon variation of $h$ of order one we expect  jumps of the total magnetization of
order $\sqrt{N}$ we define:
\begin{equation}
m_h = \frac{1}{\sqrt{N}} \,M\!\left(H + \frac{h}{\sqrt{N}}\right) = - \partial_{h} F(h)
\end{equation}
where from now on we denote $F(h)$ the free energy in the external field $H + \frac{h}{\sqrt{N}}$. Note that
$m_h$ is the sum of a constant part of order $\sqrt{N}$, $\overline{m_0}=\frac{\overline{M(H)}}{\sqrt{N}}$, plus a fluctuating
part $m_h - \overline{m_0}$ of order unity. 

To characterize the statistics of these order-one jumps in $m_h$ we need to compute the following
cumulants in different {\em physical} fields $h_{i}$, $i=1,\ldots ,p$:
\begin{eqnarray}
\overline{m_{h_1}\dots m_{h_p}}^c  =  
\partial_{h_1} \dots \partial_{h_p} S^{(p)} (h_{1},h_{2}, \dots ,h_{p})\ .
\end{eqnarray}
It is obtained from the cumulants of the sample-to-sample fluctuations of the free energy,
\begin{eqnarray}
S^{(p)} (h_{1},h_{2}, \dots ,h_{p}) = (-1)^p\, \overline{F(h_1)\dots F(h_p)}^{J,c},
\end{eqnarray}
where we denote by $\overline{\rule{0mm}{1.5ex}\dots}^J$ the average over disorder and
$\overline{{\rule{0mm}{1.5ex} }\dots}^{J,c}$ its connected average. 

These can be obtained from the generating function $W[\{h^a\}] \equiv W[h]$ of $a=1,\dots,n$ replica submitted to
different fields $h^a$, 
\begin{eqnarray}
\exp\big( W[h]\big):=\overline{
\exp\Big[-\beta \sum_{a=1}^n F(h^a) \Big]
}^{J} \ .
\end{eqnarray}
Note that fields $h^a$ with replica index $a$ are denoted with {\em upper} index to distinguish it from the physical field $h_i$ with lower index. 
Hence
\begin{eqnarray}\label{m12}
 W [h]&=& \sum_{q=0}^\infty \frac{\beta^q}{q!} \sum_{a_1,\ldots a_q} S^{(q)}(h^{a_1},\ldots ,h^{a_q})
\end{eqnarray}
We now derive a formula for $W[h]$ from the saddle-point equations in the large-$N$ limit.

\subsection{Saddle-point equations}
One has:
\begin{eqnarray}\label{m2}
 \lefteqn{\rme^{ W[h]}} \nn\\&=&  \overline{\sum_{\{\sigma^{i}_{a} \}}
\exp\bigg[{\beta \sum_{ij}\sigma_{a}^{i}J_{ij}\sigma_{a}^{j} + \beta \sum_{i}\Big(H + \frac{h^a}{\sqrt{N}} \Big) \sigma^{i}_{a}} \bigg]
}^{J}  \nn \\
&=&   \sum_{\{\sigma^{i}_{a}\}} \int \prod_{a\neq b} \rmd Q_{ab} \, \prod_{i}
\exp\bigg( nN\frac{\beta^{2}J^{2}}{2}\bigg)\nn\\
&& \qquad\times \exp\bigg[ \sum_{a}\beta \left(H + \frac{h^a}{\sqrt{N}} \right) \sigma_{a}^{i}\bigg] \nonumber  \\
&& \qquad\times \exp\bigg[ {\beta^{2}J^{2}\sum_{a\neq  b}
\left(-\frac{N}{2}Q_{ab}^{2}+Q_{ab}\sigma_{a}^{i}\sigma_{b}^{i} \right) }\bigg].
\end{eqnarray}
Note that on spins $\sigma^i_a$, we put the replica-index $a$ at the bottom, and the site index $i$ at the top.
Now we define the {\em local} partition sum
\begin{align}\label{m3}
&\rme^{A (Q,h)}  \\&:= \sum_{\{\sigma_{a}\}} \exp\!
\left[\beta^{2}J^{2}\sum_{a\neq b}Q_{ab}\sigma_{a}\sigma_{b} + \beta \sum_{a} \Big(H + \frac{h^a}{\sqrt{N}} \Big) \sigma_{a} \right], \nn
\end{align}
in terms of which we can write
\begin{eqnarray}\label{m5}
 \rme^{ W[h]} &=&  \int \prod_{a\neq b} \rmd Q_{ab} \, \exp\!
\bigg[nN\frac{\beta^2J^2}{2}-\frac{N}{2}\beta^{2}J^{2}\sum_{a\neq b}Q_{ab}^{2}  \nonumber \\
&& \qquad\qquad\qquad \qquad+ N A (Q,h)\bigg].
\end{eqnarray}
In the limit of $N \to \infty$ we can perform a saddle-point evaluation. For $h^a=0$ this is the usual SK saddle-point
equation in presence of a field $H$. In the low-temperature phase  considered here, it has a set of solutions, denoted $q_{ab}^\pi=q_{\pi^{-1}(a) \pi^{-1}(b)}$. They are obtained from the standard Parisi solution $q_{ab}$ by applying a permutation $\pi\in {\cal S}(n)$ of the indices. Each saddle point $q_{ab}$ of the path integral over $Q_{ab}$ satisfies the self-consistent equation  for $a \neq b$:
\begin{equation}\label{m4}
 \left< \sigma_{a} \sigma_{b} \right>_{A(q,0)}= q_{ab} \ ,
\end{equation}
where $\left<\dots \right>_{A}$ refers to an average with action $A$ from Eq.\ (\ref{m3}). Since changes in the external
fields are of size $h^a/\sqrt{N}$, they alter the solution of the saddle-point equation
from $q=q_0$ to $q_h=q_0 + O(\frac{1}{\sqrt{N}})$. Hence we can compute the contribution to
$W[h]$ of each saddle point in perturbation theory. For a given saddle point, each contribution to $\rme^{W[h]}$ is of the form $\rme^{V[q_h,h]}$, with
\begin{equation}
V[q,h] := nN\frac{\beta^2J^2}{2}-\frac{N}{2}\beta^{2}J^{2}\sum_{a\neq b} q_{ab}^{2}+ N A [q,h]\ .
\end{equation}
The saddle-point condition satisfied for any $h$ reads 
\begin{eqnarray} \label{stat}
\partial_{q_{ab}} V[q_h,h]=0\ .
\end{eqnarray}
Using this equation,  
we obtain the following expansion in replica fields $h^a$:
\begin{eqnarray}\label{m8}
V[q_h,h] &=& V [q,0] + \sum_{a} h^a (\partial_{h^a}V)[q,0] \nn\\
&& + \frac{1}{2} \sum_{ab} h^a h^b (\partial^2_{h^a h^b}V)[q,0] + O({\textstyle \frac{1}{\sqrt{N}}})\nn \\
&=& V [q,0] + \beta  \sum_{a} h^a \overline{m_{0}} \nn \\
&& + \frac{1}{2} \sum_{ab} h^a h^b \beta^2 \Big[ \delta_{ab} + (1-\delta_{ab}) q_{ab} \Big]+ O({\textstyle \frac{1}{\sqrt{N}}}) \nn
\end{eqnarray}
In the first line we used the 
condition (\ref{stat}),  its total derivative w.r.t.\ $h^a$, and that $\partial_{h} q_{h} = O(\frac{1}{\sqrt{N}})$ to eliminate the cross term $\partial _q\partial_h V$; in the second line we used Eq.\ (\ref{m4}). 

The final expression for $W[h]$ is obtained by performing the sum over all saddle points $q^{\pi}_{ab}=q_{\pi^{-1}(a) \pi^{-1}(b)}$,
\begin{eqnarray}\label{m5bis}
&& \rme^{ W[h] - W[0] - \beta  \sum_{a} h^a \overline{m_{0}} } \nn \\
&& \qquad =
\sum^\prime_\pi  e^{ \frac{\beta^2}{2} \sum_{a} (h^a)^2 (1- q_{aa}) + \frac{\beta^2}{2} \sum_{ab}
 q^\pi_{ab} h^a h^b} \ .\qquad
\end{eqnarray}
The prime on the permutation sum indicates that the sum is normalized by $\sum^\prime_\pi 1=1$. For convenience, we have introduced  $q_{aa}$ 
to be defined later. 

Let us define the ``non-trivial" part $\tilde W[h]$ of $W[h]$ as
\begin{eqnarray}
\tilde W[h]&:=&W[h] - W[0] - \beta  \sum_{a} h^a \overline{m_{0}} \nn\\&& 
- \frac{\beta^2}{2} \sum_{a} (h^a)^2 (1- q_{aa}) 
\nonumber  \\
&=&  
\ln \left( \sum^\prime_\pi  e^{ \frac{\beta^2}{2} \sum_{ab} q_{ab} h^{\pi(a)} h^{\pi(b)}}\right).
\end{eqnarray}
To obtain the $p$-th cumulant, we need to consider $W[\{h^a\}]$ for $p$ groups of $n_1=\alpha_1 n,n_2=\alpha_2 n,\ldots ,n_p=\alpha_p n$ replica with $\sum_{i=1}^p \alpha_i=1$.  Each group is subject to a different physical field $h_i$, $i=1,\ldots ,p$. This field is constant within a replica group. We remind that we use superscript indices $h^{a} $ to denote replicas, and subscript indices $h_{i}$ to label the replica groups. The resulting $ W_p[h]:=W(h_1,\dots,h_p):= W[\{h^a\}] $ (and likewise for $\tilde W_p[h]$) has the cumulant expansion
\begin{equation} \label{whh}
W_p[h] = \sum_q \frac{\beta^q}{q!} n^q
 \sum_{i_1=1}^p\ldots \sum_{i_q=1}^p \alpha_{i_1}\ldots  \alpha_{i_q} S^{(q)}(h_{i_1},\ldots h_{i_q}) \, .
\end{equation}
The magnetization cumulants for $p>1$ can then be extracted as
\begin{eqnarray}
&& \overline{m_{h_1}\ldots m_{h_p}}^{J,c}= \partial_{h_1}\ldots \partial_{h_p} 
S^{(p)}(h_1,\ldots ,h_p) \nonumber \\
&&  \label{cum1}
=\lim_{n\to 0}\frac{1}{n^p\beta^p\prod_{i=1}^p \alpha_i} \partial_{h_1} \ldots  \partial_{h_p} \tilde W_{p}[h]\ .
\end{eqnarray}
This works, 
since the terms in (\ref{whh}) with $q<p$ vanish after the differentiation and the ones with $q>p$
vanish in the limit $n \to 0$, leaving the desired term $q=p$.

\section{Calculation of moments}
\subsection{Generalized flow equation}
To proceed, we  decouple the $h_a$'s by a Hubbard-Stratonovich transformation, 
\begin{eqnarray}
\label{tobecomp}
\rme^{\tilde W[h]}=
\la \sum_{\pi}^\prime \rme^{ \sum_{a} h^{\pi(a)} \mu_{a}} \ra_{\mu},
\end{eqnarray}
where $\mu_a$ are Gaussian random variables with variance $\la \mu_a\mu_b\ra_{\mu}= \beta^2 q_{ab}$, and $\la\ra_\mu$ denotes the average over them.

\begin{figure}
\centerline{\includegraphics[width=0.45\textwidth]{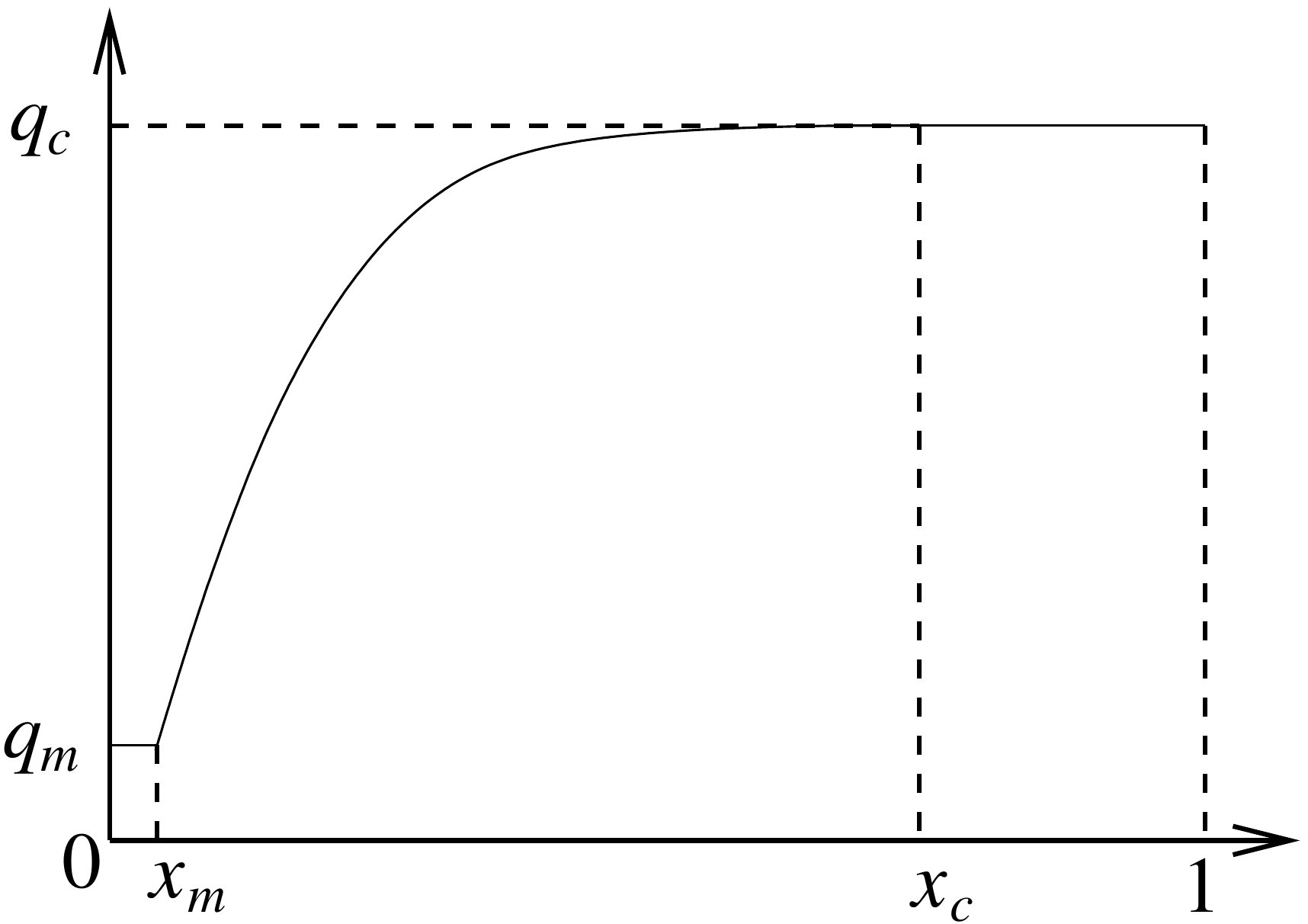}}
\caption{The Parisi-function $q(x)$, with its two plateaus for $x<x_{m}$ and $x>x_{c}$. Note that this gives two $\delta$-function contributions to the derivative of the inverse function, $\frac{\rmd x(q)}{\rmd q} = x_{m} \delta(q-q_{m}) + (1-x_{c}) \delta(q-q_{c}) + \mbox{smooth part}$. }
\label{f:parisi}
\end{figure}

The sum over permutations in (\ref{tobecomp}) is equivalent to a normalized sum (indicated by a prime) over assignments $\{i_a\}\in \{1,\dots, p\}$, describing the permutation $\pi$:
\begin{eqnarray}
h^{\pi(a)}= h_{i_a}.
\end{eqnarray}
Since the permutation preserves the number of equivalent replica, we have the constraint $\sum_a\delta_{j,i_a}=n\alpha_j$.
With this notation we obtain
\begin{equation}
\label{HStrafo}
\rme^{\tilde W_p[h]}=
\la \sum_{i_a\in\{1,\dots,p\} |\sum_a\delta_{j,i_a}=n\alpha_j}^\prime \exp\left(\sum_{a} 
h_{i_a}\mu_a\right)\ra_{\mu}.
\end{equation}
As we prove in App.~\ref{app:proof}, this can be rewritten as
\begin{widetext}
\begin{eqnarray}
\label{Theorem}
\rme^{\tilde W_p[h]}=
\la \frac{\int_{-\infty}^\infty \prod_{i=1}^p \rmd y_i \, \delta(\sum_{i=1}^p \alpha_iy_i) \prod_{a=1}^n \left[\sum_{i=1}^p\exp( h_{i}\mu_a+y_i)\right]}{\int_{-\infty}^\infty \prod_{i=1}^p \rmd y_i \, \delta(\sum_{i=1}^p \alpha_iy_i) \left[\sum_{i=1}^p\exp(y_i)\right]^{n}}\ra_{\mu},
\end{eqnarray}
\end{widetext}
valid for any $n<0$, and for any {\red set of $\alpha_i>0$, with $\sum_{i=1}^p \alpha_i=1$.}
This identity significantly generalizes the formula (D6) in Ref.~\onlinecite{BMP}.

In the case where $q_{ab}$ is a hierarchical ultrametric matrix of Parisi type,
parameterized by the Parisi function $q(x)$ with $n<x<1$, 
the average over $\mu_a$ of expression (\ref{Theorem}) can be
performed extending the methods of Ref.~\onlinecite{Duplantier}. We recall that we use everywhere
$\sum_i \alpha_i=1$ and rewrite
\begin{equation}
\label{Wnt}
\rme^{\tilde W_p[h]}= \frac{\int_{-\infty}^\infty \prod_{i=1}^p \rmd y_i \,
\delta(\sum_i\alpha_i y_i)\,
g(x=n,\{y_i\})}{\int_{-\infty}^\infty
\prod_{i=1}^p \rmd y_i \, \delta(\sum_i\alpha_i y_i)\left[\sum_i
\exp(y_i)\right]^{n}}\, .
\end{equation}
We have defined
\begin{eqnarray}
g\left(x;\{y_i\}\right)&\equiv& \rme^{x\phi(x;\{y_i\})}\nn\\ &\equiv&
\left\langle{\prod_{a=1}^x\left(\sum_{i=1}^p e^{y_i+
h_{i}\mu_a^{(x)}}\right)}\right\rangle_{\mu^{(x)}} 
\end{eqnarray}
The
auxiliary fields $\mu_a^{(x)}$ have Gaussian correlations $\left\langle
\mu_a^{(x)} \mu_b^{(x)}\right\rangle_{\mu^{(x)}}= \beta^2[q_{ab}-q(x)]$.
For convenience,    we define $q_{aa}=q(1)$. 

Generalizing the method of Ref.\ \onlinecite{Duplantier} to several groups, we find the flow equation for 
the function $\phi(x,\{y_i\})$ defined above:
\begin{equation}
\label{Duplantierflow}
\frac{\p \phi}{\p
x}=-\frac{\beta^2}{2}\sum_{i,j=1}^p  h_i h_j \frac{\rmd q(x)}{\rmd x}\left(\frac{\p^2\phi}{\p
y_i\p y_j}+x\frac{\p \phi}{\p y_i} \frac{\p \phi}{\p y_j}\right) \ .
\end{equation}
It must be solved with the boundary condition
\begin{equation}\label{H}
\phi(x=1;\{y_i\})= \log\left(\sum_{i=1}^p \rme^{{y_i}}\right)\equiv H(\{y_i\})\ .
\end{equation}
Here and below we denote $\vec y \equiv \{y_i\}$. 

To simplify (\ref{Wnt}) we first evaluate the denominator. In the limit of $n \to 0$ one 
can show the general formula for any $\alpha_i$ with constraint $\sum_i \alpha_i=1$ and $n<0$:
\begin{eqnarray}
\label{normaliz}
&& \int \rmd^p y\, \delta\Big(\sum_i \alpha_i y_i\Big) \rme^{n H(\vec y)} \nn\\
&&= \int \rmd^p y \,\delta \Big(\sum_i \alpha_i y_i \Big) \rme^{- (-n) {\rm max}(y_{i}) } \left[1+O(n)\right] \nn\\
&&= \frac{1}{\prod_i\alpha_i} (-n)^{1-p} \left[1+O(n)\right] .
\end{eqnarray}
Expanding also the numerator and the exponential in (\ref{Wnt}) to lowest non-trivial order in $n$, we find
\begin{eqnarray}
\label{Wnt_der}
\lefteqn{\partial_{h_1} \ldots  \partial_{h_p} \tilde W_p[h]}\nn\\
&=&  \frac{n \int \rmd^p y \, \delta(\sum_i \alpha_i y_i)\partial_{h_1} \ldots  \partial_{h_p} \phi (0,\vec y)}
{(-n)^{1-p}/\prod_i\alpha_i}
\left[ 1+ {O}(n) \right] \nn\\
&=&  -(-n)^{p} \prod_{i=1}^p \alpha_i   \int_{-\infty}^\infty \rmd^py \,
\delta \Big(\sum_i \alpha_i y_i \Big)
  \partial_{h_1} \ldots  \partial_{h_p}\phi(0, \vec y) \nonumber \\
  && \times \left[ 1+ {O}(n) \right] .
\end{eqnarray}
Inserting this into Eq.~(\ref{cum1}) we obtain the final formula for the $p$-th cumulant
of the reduced magnetization:
\begin{eqnarray}
\lefteqn{\overline{m_{ h_1} \ldots  m_{h_p}}^{J,c}}\nn\\
&&=   -(-T)^{p} 
\int \rmd^py\, \delta \Big(\sum_i \alpha_i y_i \Big) \partial_{h_1} \ldots  \partial_{h_p} \phi (0, \vec y) . \qquad \label{cumulantM2}
\end{eqnarray}
This expression is independent of the choice of $\alpha_i$, as it must be. In case the order parameter function, $q(x)$, has a plateau for $x<x_m$ (as happens for the SK model in a magnetic field $H\neq 0$)
\begin{eqnarray}
\phi(x=0,\vec{y})= \overline{ \phi(x_m,\vec{y}+z\beta \vec{h} \sqrt{q_0} )}^z,
\end{eqnarray}
where $\overline{\cdots }^z$ denotes an average over $z$, a unit-centered Gaussian variable, 
\begin{equation}
\label{33}
\overline{ f(z)}^z := \int_{-\infty}^{\infty} \frac{\rmd z}{\sqrt{2 \pi}} \rme^{-z^2/2} f(z) \ ,
\end{equation}
see Eq.~(96) in Ref.~\onlinecite{FRGRSB}.

\subsection{TBL-shock expansion}
We now solve the flow equation (\ref{Duplantierflow}) perturbatively in the nonlinear term. This generates a low-temperature expansion which is well suited to study shocks\cite{FRGRSB}.
We write
\begin{eqnarray}
\phi(x,{\vec y})=\phi^0(x,{\vec y})+\phi^1(x,{\vec y})+\dots \ .
\end{eqnarray}
The successive terms satisfy
\begin{eqnarray}
\label{Duplantier0_1}
\frac{\p  \phi^0}{\p x}&=& -\frac{\beta^2}{2}\sum_{ij} h_i h_j \frac{\d q(x)}{\d x}\, \frac{\p^2\phi^0}{\p y_i\p y_j}
\end{eqnarray}
with initial condition $\phi^0(x=1,{\vec y})=H({\vec y})$.
\begin{equation}
\label{Duplantier1_1}
\frac{\p \phi^1}{\p x}= -\frac{\beta^2}{2}\sum_{ij} h_i h_j \frac{\d q(x)}{\d x}\left( \frac{\p^2 \phi^1}{\p y_i\p y_j}+x \frac{\p \phi^0}{\p y_i}\frac{\p \phi^0}{\p y_j}\right), 
\end{equation}
with initial condition $\phi^1(x=1,{\vec y})=0$. 

The  leading-order equation (\ref{Duplantier0_1}) is a linear diffusion equation, and integrated as (for $x\geq x_m$)
\begin{eqnarray}\label{32}
\phi^0(x,\vec{y})&=&\overline {H(\vec{y}+z \beta \vec h \sqrt{q(1)-q(x)})}^z.
\end{eqnarray}
Taking into account (\ref{33}), we find the contribution of $\phi^0$ to the magnetization cumulants
\begin{eqnarray}
\lefteqn{\overline{m_{ h_1} \ldots  m_{h_p}}^{J,c,(0)}}\nn\\
&=&   -(-T)^{p} 
\int \rmd^py\, \delta \Big(\sum_i \alpha_i y_i \Big) \nn\\
&& \qquad \times \partial_{h_1} \ldots  \partial_{h_p} \overline {H\!\Big(\vec{y}+z \beta \vec h \sqrt{q(1)}\Big)}^z . \qquad \ \ \ \label{c3}
\end{eqnarray}
It is shown in  appendix \ref{a:A} that at $T=0$  this equals 
\begin{eqnarray}
\lefteqn{\overline{m_{ h_1} \ldots  m_{h_p}}^{J,c,(0)}}\nn\\
&=&q(1)^{p/2} \, \overline{z^p} = \left[2q(1)\right]^{p/2} \frac{ \left[(-1)^p+1\right] \Gamma \left(\frac{p+1}{2}\right)}{2\sqrt{\pi }}, \qquad 
\end{eqnarray}
which is a constant independent of $h_i$. In addition $q(1)\to 1$ as $T\to 0$. For $p=2$ one finds at any temperature the contribution
\begin{equation}
\label{mm0}
\overline{m_{ h_1}^2}^{J,c,(0)} = q(1).
\end{equation}
Even though at $T=0$ this is the full result for the sample-to-sample fluctuations of the magnetization, at finite $T$ there will be an additional piece from $\phi^1$ obtained below.  Similarly, to obtain the full finite-$T$ expression of higher-order moments of $m_{h_1}$, contributions from $\phi^{p>0}$ are needed. However, here we focus on $T=0$. 

We now turn to the calculation of the contributions which capture the information about jumps, which are of order ${\cal O}(|h_i-h_j|)$ in the limit $T\to 0$. It is contained in the contribution of $\phi^1$ and only in that contribution, as was discussed in Ref.~\onlinecite{FRGRSB}. Higher-order functions $\phi^p$ contain contributions of order ${\cal O}(|h_i-h_j|^p)$ at $T=0,$ encoding information of multi-shock correlations. 
To first order in the non-linear term we find, extending the calculation in Ref.~\onlinecite{FRGRSB}: 
\begin{widetext}
\begin{eqnarray}
\phi^1(x,\vec{y})&=&\int_x^1 \rmd x'\, \frac{\beta^2}{2}\sum_{ij}\frac{\d q(x')}{\d x'} x' h_i h_j \overline{\frac{\p \phi^0}{\p y_i}\left(x',\vec{y}+\eta\,\beta \vec{h} D_{x'x}\right)\frac{\p \phi^0}{\p y_j}\left(x',\vec{y}+\eta\,\beta \vec{h}D_{x'x}\right)}^\eta\nn\\
&=&\int_x^1 \rmd x'\, \frac{\beta^2}{2}\sum_{ij}\frac{\d q(x')}{\d x'} x' h_i h_j \overline{\frac{\p H}{\p y_i}\left(\vec{y}+\beta\vec{h}\left[\eta D_{x'x}+z_1 D_{1x'}\right]\right)\frac{\p H}{\p y_j}\left(\vec{y}+\beta \vec{h}\left[\eta D_{x'x}+z_2 D_{1x'}\right]\right)}^{\eta,z_1,z_2}.\qquad
\end{eqnarray}
\end{widetext}
As in  Eq.~(\ref{33}), $\eta$, $z_1$ and $z_2$ are independent unit-centered Gaussian random variables, and $D_{x'x}:=\sqrt{q(x')-q(x)}$. 

We now change integration variables from $x\to q$ and define $\hat x(q):= x(q)/T$ and $\hat h:=h/T$, the ``thermal boundary layer variable"  \cite{FRGRSB} for the external field. Using Eq.~(\ref{cumulantM2})  the contribution of the first-order term to the magnetization cumulant becomes, denoting $q_c:=q(x_c)$, and $q_m:=q(0)$, see Fig.~\ref{f:parisi},
\begin{widetext}
\begin{eqnarray}
\lefteqn{ \overline{m_{h_1} \ldots  m_{h_p}}^{J,c,(1)}} \nn\\
&=&  (-1)^{p+1}
 \partial_{\hat h_1} \ldots  \partial_{\hat h_p}  \frac{T}{2 }  \int_{q_m}^{q_c} \rmd q\, \hat x(q)  \int_{-\infty}^\infty \prod_{i=1}^p \rmd y_i\,  
\delta\Big(\sum_i \alpha_i y_i\Big)  \overline{\partial_{A_+}\partial_{A_-} H\Big(\vec{y}+\vec{\hat h}A_+\Big)H\Big(\vec{y}+ \vec{\hat h}A_-\Big)}^{A_+,A_-} \nn\\
&=&  (-1)^{p}
 \partial_{\hat h_1} \ldots  \partial_{\hat h_p}  \frac{T}{2 }  \int_{q_m}^{q_c} \rmd q\, \hat x(q)  \int_{-\infty}^\infty \prod_{i=1}^p \rmd y_i\,  
\delta\Big(\sum_i \alpha_i y_i\Big)  \overline{\partial_{A_+}\partial_{A_-} \frac12\left[ H\Big(\vec{y}+\vec{\hat h}A_+\Big)-H\Big(\vec{y}+ \vec{\hat h}A_-\Big) \right]^2}^{A_+,A_-}\ .\qquad
\end{eqnarray} 
\end{widetext}
$A_\pm$ are centered Gaussian random variables with correlations defined from the above independent Gaussian variables as
\begin{eqnarray}
&& A_{+}=\eta \sqrt{q-q_m} +z_1 \sqrt{q_c-q} \\
&&  A_{-}=\eta \sqrt{q-q_m} +z_2 \sqrt{q_c-q}\ .
\end{eqnarray}
It is convenient to introduce
\begin{eqnarray}
F:=A_+ + A_-\ , \qquad G:=A_+-A_- 
\end{eqnarray}
in terms of which one can integrate by part
\begin{widetext}
\begin{eqnarray}
\label{TBLmoments}
 \overline{m_{h_1} \ldots  m_{h_p}}^{J,c,(1)}&=&  (-1)^{p}
 \partial_{\hat h_1} \ldots  \partial_{\hat h_p}  \frac{T}{2 }  \int\limits_{q_m}^{q_c} \rmd q\, \hat x(q)  
 \int\limits_{-\infty}^{-\infty} \rmd F \int\limits_{-\infty}^{-\infty}\rmd G\, (\partial^2_{F}-\partial^2_{G}) \frac{\exp\!\big(-\frac{F^2}{4[q_c+q-2q_m]}-\frac{G^2}{4[q_c-q]}\big)}{2\pi \sqrt{2[q_c+q-2q_m]2(q_c-q)}}  \nonumber \\
 && \times 
 \prod_{i=1}^p \int_{-\infty}^\infty \rmd y_i  \,
\delta\Big(\sum_i \alpha_i y_i\Big)\frac{1}{2} \left[H\left(\vec{y}+\half \vec{\hat h}(F+G)\right)-H\left(\vec{y}+ \half \vec{\hat h}(F-G) \right)\right]^2 \ .\qquad 
\end{eqnarray}
The differential operator $\partial_{A_+}\partial_{A_-} = \partial^2_{F}-\partial^2_{G}$ acts only on the Gaussian measure. Note that its action is equivalent to $ \partial^2_{F}-\partial^2_{G} \equiv \rmd/\rmd q$. One can thus integrate by part over $q$ to get
\begin{eqnarray}
\label{TBLmoments-2}
 \overline{m_{h_1} \ldots  m_{h_p}}^{J,c,(1)}&=& \frac{ (-1)^{p+1} T}{4 }
 \partial_{\hat h_1} \ldots  \partial_{\hat h_p}    \int\limits_{q_m}^{q_c} \rmd q\, \frac{\rmd \hat x(q)}{\rmd q}  
 \nonumber \\
 && \times 
 \prod_{i=1}^p \int_{-\infty}^\infty \rmd y_i  \,
\delta\Big(\sum_i \alpha_i y_i\Big) \overline{\left[H\left(\vec{y}+\half \vec{\hat h}(F+G)\right)-H\left(\vec{y}+ \half \vec{\hat h}(F-G) \right)\right]^2}^{F,G} \ .\qquad 
\end{eqnarray}
The measure over $F$ and $G$ is defined by
\begin{equation}\label{measureFG}
\overline{f(F,G)}^{F,G}:= \int\limits_{-\infty}^{-\infty} \rmd F \int\limits_{-\infty}^{-\infty}\rmd G\, \frac{\exp\!\Big(-\frac{F^2}{4[q_c+q-2q_m]}-\frac{G^2}{4[q_c-q]}\Big)}{2\pi \sqrt{2[q_c+q-2q_m]2(q_c-q)}} f(F,G).
\end{equation}
Equivalently, one can write in terms of $A_+$ and $A_-$
\begin{equation} \label{startingpoint}
 \overline{m_{h_1} \ldots  m_{h_p}}^{J,c,(1)}=  \frac{(-1)^{p+1} T }{4}
\partial_{\hat h_1} \ldots  \partial_{\hat h_p}  \int\limits_{q_m}^{q_c} \rmd q\, \frac{\rmd\hat x(q)}{\rmd q}  \prod_{i=1}^p \int\limits_{-\infty}^\infty  \rmd y_i\,  
\delta\Big(\sum_i \alpha_i y_i\Big)  \overline{ \left[H\left(\vec{y}+\vec{\hat h}A_+\right)-H\left(\vec{y}+ \vec{\hat h}A_-\right)\right]^2 }^{A_+,A-}
\end{equation}
with measure 
\begin{equation}
\overline{f(A_+,A_-)}^{A_+,A_-}:= \int\limits_{-\infty}^{-\infty} \rmd A_+ \int\limits_{-\infty}^{-\infty}\rmd A_-\, \frac{\exp\!\Big(-\frac{(A_++A_-)^2}{4[q_c+q-2q_m]}-\frac{(A_+-A_-)^2}{4[q_c-q]}\Big)}{\pi \sqrt{2[q_c+q-2q_m]2(q_c-q)}} f(A_+,A_-).
\end{equation}
\end{widetext}
The boundary terms in the integration by part vanish, provided that whenever $q(x)$ exhibits a plateau for $0 \leq x \leq x_m$ it is included as a $\delta$ function. 

Using the expression (\ref{H}) for $H(y)$, the formula (\ref{TBLmoments-2}) allows us to compute the thermal boundary-layer form of the $p$-th cumulant. We give here the result for $p=2$:
\begin{eqnarray}
\label{2ndcumulant}
 \lefteqn{ \overline{m_{h_1} m_{h_2}}^{J,c,(1)} }\nn\\
 &=& - \frac{1}{4}  \int_{q_m}^{q_c} \rmd q\, \frac{\rmd\hat x(q)}{\rmd q}  
\int_{-\infty}^\infty \rmd G\, \Bigg[ \frac{\exp\left(-\frac{G^2}{4[q_c-q]}\right)}{ \sqrt{4\pi [q_c-q]}} \Bigg] \nn\\
&& \qquad \qquad\times  G^3 ( h_1- h_2) \coth\!\left(\frac{(h_1- h_2) G}{2T}\right),
\end{eqnarray}
recovering  the form obtained in Ref \cite{FRGRSB}. For $T>0$ and $h_2\to h_1$ one finds
\begin{eqnarray}
 \overline{m_{h_1} m_{h_1}}^{J,c,(1)} = \int_0^1 q(x) \,\rmd x -q(1).
\end{eqnarray}
Added to Eq.~(\ref{mm0}), this gives the correct total sample-to-sample fluctuations of the magnetization.
The fact that higher terms $\phi^p$ do not contribute to this variance can be verified by a direct expansion of Eq.~(\ref{Duplantierflow}) in $q(x)$.

For general $p$ we only study the limit $T \to 0$. For convenience we introduce the
notation $A_M\equiv {\rm max}(A_+,A_-)=(F+|G|)/2$, $A_m\equiv {\rm min}(A_+,A_-)=(F-|G|)/2$.
The calculation is performed in Appendix \ref{s:app1} and we obtain  for $p \geq 2$:
\begin{widetext}
\begin{equation} \label{overlined} 
  \overline{m_{h_1} \ldots  m_{h_p}}^{J,c,(1)}=  
   \frac{1}{2 }  \int_{q_m}^{q_c} \rmd q\, \frac{\rmd\hat x(q)}{\rmd q}    \overline{  (A_M-A_m)\left(- h_p A_m^p +\sum_{m=1}^{p-1} (h_{p-m+1}-h_{p-m}) A_m^{p-m}A_M^{m}  +h_1 A_M^p \right) }^{A_+,A-}.
\end{equation}
Note that we have put back the physical field $h = T\hat h$, making evident the result in the limit of $T\to 0$. 
\end{widetext}
As an example, for $p=2$ we obtain
\begin{eqnarray} \label{overlined2} 
  \overline{m_{h_1} m_{h_2}}^{J,c,(1)}=  -
   \frac{1}{4 }  \int_{q_m}^{q_c} \rmd q\, |h_1-h_2|\frac{\rmd\hat x(q)}{\rmd q} \,   \overline{ |G|^3 }^{G}\ , 
\end{eqnarray}
which is the $T=0$ limit of (\ref{2ndcumulant}). Note that this describes the correction to order $|h_2-h_1|$ to the 2-point function of the magnetization (\ref{mm0}).
This encodes the second moment of the jump-size distribution, as was discussed in Ref.~\onlinecite{FRGRSB} for the random-manifold problem.
We now turn to the determination of the full distribution from the above cumulants.

\subsection{Distribution of jumps}
We now derive the distribution of jumps  by showing that the above result is identical to a $p$-point correlator of the magnetization of a  two-level system, whose characteristics (jump size and jump location) are distributed in a simple manner. We notice that
the above expression (\ref{overlined}) for the cumulants can be reexpressed as
\begin{eqnarray} \label{super}
\lefteqn{ \overline{m_{h_1} \ldots  m_{h_p}}^{J,c,(1)}} \\
&=&   \frac{1}{2 }  \int_{q_m}^{q_c} \rmd q\, \frac{\rmd\hat x(q)}{\rmd q}   \nn\\
   && \times \overline{ |G| \left\langle  
    \mu(h_1)\cdots \mu(h_p)-\mu(0)\cdots \mu(0) \right\rangle_{h_c} }^{F,G}\nn
\end{eqnarray}
in terms of the  ``random magnetization" variable
\begin{eqnarray}
\mu({h})&=&\theta({h}-{h}_c)A_M +\theta({h}_c-{h})A_m
\nn\\
&=& \frac F2 +{\rm sign}(h-h_c)\frac{|G|}{2}\ .
\end{eqnarray}
It exhibits a jump of size $|G|$ at location $h=h_c$, uniformly distributed on the real axis with unit density.
$F/2$ is interpreted as the mean magnetization. 
To prove this, we have to show that 
\begin{eqnarray}
\label{jumpmoments}
&&\!\!\!{- h_p A_m^p +\sum_{m=1}^{p-1} (h_{p-m+1}-h_{p-m}) A_m^{p-m}A_M^{m}  +h_1 A_M^p} \nonumber \\
&& \qquad =  \left\langle  
    \mu(h_1)\cdots \mu(h_p)-\mu(0)\cdots \mu(0) \right\rangle_{h_c} \ ,\qquad
\end{eqnarray}\nopagebreak
where the overbar denotes averaging with respect to $h_c$.
To see that, consider an interval $[{h}_0,{h}_L]$ containing $0$ and  ${h}_0<0<{h}_1<...<{h}_p<{h}_L$, in which  $h_c$ is uniformly distributed with density $\rho_0=1$. 

The probability of ${h}_c$ to be in the interval $[{h}_{p-m+1},{h}_{p-m}]$ is $\rho_0 ({h}_{p-m+1}-{h}_{p-m})$.
Then $\overline{\mu({h}_1)\cdots \mu({h}_p)}^{h_c}$ gives the corresponding term in the sum  (\ref{jumpmoments}) (multiplied by $\rho_0$).  At the edge, when $h_c<h_1$, it gives $\rho_0 ({h}_1-{h}_0) A_M^p $. When $h_c>h_p$ it gives $ \rho_0 ({h}_L-{h}_p) A_m^p$. Subtracting $\overline{\mu(0)\cdots \mu(0)}=\rho_0 {h}_L A_m^p -
\rho_0 {h}_0 A_M^p$  yields (\ref{jumpmoments}), (multiplied by $\rho_0$, which is set to unity).
Both mean magnetization $F/2$ and jump-size $|G|$ are obtained from Gaussian
variables with a $q$-dependent variance. The full result in (\ref{super}) is obtained by weighting with the probability distribution of the various values of $q$. 

A given shock at field $h=h_s$ is characterized by its magnetization jump of size $\Delta m = m_{h^+}-m_{h^-}$ (always $>0$), and its mean magnetization $m_s= \frac{1}{2}\left[M(H+\frac{h}{\sqrt{N}})+M(H)\right]$ at the shock. From the above we can  extract the joint density (per unit interval of $h$), $\rho( \Delta m,\delta m)$, of shocks of size $\Delta m$, and shift $\delta m = m_s-\overline{m_0}$. It is   defined as:
\begin{widetext}
\begin{equation}
\label{shockdistribution}
\rho(\Delta m,\delta m)=\lim_{ h\downarrow 0} \frac{1}{ h}\,\overline{ \delta\!\!\left(\Delta m- \frac{M(H+\frac{h}{\sqrt{N}})-M(H)}{\sqrt{N}}\right) \delta\left(\delta m -\frac{M(H+\frac{h}{\sqrt{N}})+M(H)-2\overline{M(H)}}{2\sqrt{N}}\right)\!},
\end{equation}
and can be extracted from Eqs.~(\ref{super}) and (\ref{measureFG}), identifying $\delta m$ with $F/2$, and $\Delta m $ with $|G|$ as discussed above. This leads to  
\be
\rho( \Delta m,\delta m) = \theta(\Delta m) \Delta m
 \int_{q_m}^{q_c} \rmd q\,  \frac{\rmd\hat x(q)}{\rmd q}  \frac{\exp\!\Big(-\frac{(\delta m)^2}{[q_c+q-2q_m]}\Big)}{ \sqrt{\pi[q_c+q-2q_m]}} \frac{\exp\!\Big(-\frac{(\Delta m)^2}{4[q_c-q]}\Big)}{\sqrt{4\pi [q_c-q]}} .
\ee
\end{widetext}
Note that after integration over $q$, the jump size and the magnetization shift become correlated. 
Integrating out the magnetization shift, we obtain our main result, the density of shock sizes per unit interval of $h$ (at $T=0$):
\begin{equation}
\label{rhoSK}
\rho( \Delta m) = \theta(\Delta m) \Delta m  \int_{q_m}^{q_c}\! \rmd q\,  \frac{\rmd\hat x(q)}{\rmd q}  \frac{\exp\!\Big(\!-\frac{(\Delta m)^2}{4[q_c-q]}\Big)}{\sqrt{4\pi [q_c-q]}} .
\end{equation}
This formula is valid for a large class of models described by replica symmetry breaking saddle points, as emphasized in \cite{us}. Here, we will focus on its application to the SK model.
Apart from the prefactor $\Delta m$, the above formula is essentially a superposition of Gaussians (at fixed overlap distance $q$). Their contribution is weighted by the density 
\begin{equation}
\label{nu_q}
\nu(q)  =\frac{\rmd \hat x(q)}{\rmd q} = \frac{1}{T} P(q).
\end{equation}
where $P(q)$ is the sample averaged probability distribution of overlaps between metastable states sampled from the Gibbs distribution.~\cite{MPV}
The weight $\nu(q)$ can be interpreted as the probability density of finding a metastable state at overlap within $[q,q+\rmd q]$ and energy within $[E,E+\rmd E]$, with $E$ close to the ground state. We will come back to this interpretation below.


A useful check of Eq.~(\ref{rhoSK}) is provided by the average magnetization jump. It is
\begin{eqnarray}
&&\!\!\!{\int_0^\infty  \rho(\Delta m) \Delta m \, \rmd\Delta m =
\int_{q_m}^{q(x_{c})} \rmd q \frac{\rmd \hat x(q)}{\rmd q}[q_c-q( x)]} \nn\\
&&= \lim_{T\to 0} \frac{1}{T}\int_0^1 \rmd x \,[q_c-q( x)] \ ,
\end{eqnarray}
where we remind that our definition of $\hat x(q)$ contains a $\delta$-function contribution  at each plateau, so that the final integral over $x$ runs again from 0 to 1.  
This formula is generally valid~\cite{MPV}.  For the SK model it can be rewritten in terms of the thermodynamic (field cooled) susceptibility, 
\begin{eqnarray}
&&\int_0^\infty  \rho(\Delta m) \Delta m \, \rmd\Delta m = \lim_{T\to 0} 
[\chi_{\rm FC}-\chi_{\rm ZFC}] = \chi_{\rm FC}^{(T=0)},\nn
\end{eqnarray}
since in the SK model the intra-state (zero-field cooled) susceptibility, $\chi_{\rm ZFC} = [1-q_c]/T$, vanishes linearly as $T\to 0$. Thus, the response is entirely due to inter-state transitions in the form of avalanches (shocks).  This is in contrast to other mean-field models, where even at $T=0$ part of the response is due to smooth intra-state polarizability\cite{us}.

\section{Application to the SK model}

\subsection{Study of the distribution of jumps: $H=0$} 
\begin{figure}
\centerline{\includegraphics[width=0.45\textwidth]{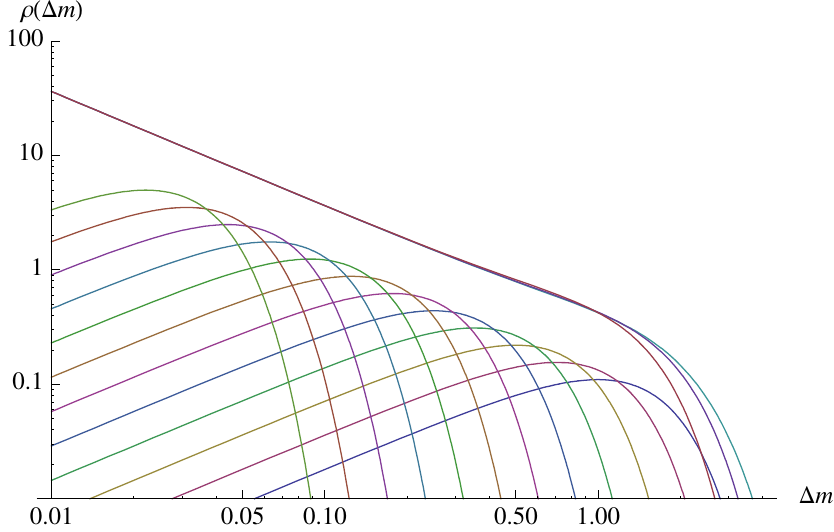}}
\caption{Power-law density of jump sizes for the SK model. The power law receives contributions from all overlaps $1-q$. The curves in the lower part show the contributions
from $(1-q)=2^{-k}$, $k=1,...,12$, each of which takes the form of the jump density in mean field glasses with 1-step replica symmetry breaking. The three nearly coinciding lines on the top show $\rho(\Delta m)$ evaluated from Eq.~(\ref{rhoSK}), for external fields $H= 0,0.25$ and $0.5$, respectively, using approximations for $q(\hat x)$ described in the text. 
The increase of $H$ decreases the cutoff at large $\Delta m$, while the avalanche distribution for $\Delta m\ll 1$ is a universal power law, not affected by $H$.}
\label{fig:illustration}
\end{figure}

In order to evaluate the distribution of jumps, we need the full replica-symmetry breaking solution of the SK model in the limit of $T \to 0$. The increasing function $q(x)$ is well characterized~\cite{ParisiToulouse,Pankov06,MuellerPankov07,opperman}, even though no closed analytical formula is known. $q(x)$ has a continuous part up to the ``break point'' $x_c\approx 0.55$, and is constant for $x_c\leq x\leq 1$. In the limit of $T\to 0$ this constant $q_c$ behaves as $1-1.592 T^2$, and $q(x)$  
 becomes essentially a function of $\hat x = x/T$ that we call $q(\hat x)$. In the absence of a magnetic field $H=0$, $q_m=0$ and $q(\hat x)\approx \frac{\hat x}{\nu(0)}$  with $\nu(0)= 1.34523$ at small $\hat x$. ~\cite{opperman} At large $\hat x$ the function crosses over to the asymptotic behavior $1-q(\hat x)\approx 4C^2/\hat x^2+ B T^2$ with $C= 0.32047$ and $B=O(1)$. This leads to  
a power-law tail for the weight of large overlaps $q\to 1$~\cite{Pankov06},
\begin{equation} 
\label{qSK}
\nu(q|1\gg 1-q\gg T^2 ) =
C (1-q)^{-3/2}\ .
\end{equation} 
We can now analyze the jump-size  density using formula (\ref{rhoSK}). We obtain analytical expressions in the  limits of small and large $\Delta m$. Numerical calculations describing the full range are shown in Fig.~\ref{fig:illustration}.
For small $\Delta m$ the integral over $q$ is controlled by $1-q\ll 1$, and we can approximate
\begin{eqnarray} \label{rhoDeltam}
\rho(\Delta m) &\approx & \int_{-\infty}^1 \frac{C \rmd q }{(1-q)^{3/2}} \Delta m \frac{\exp\!\Big(-\frac{(\Delta m)^2}{4(1-q)}\Big)}{\sqrt{4\pi (1-q)}} \nn\\
&=&\frac{2C}{\sqrt{\pi}} \frac{1}{(\Delta m)^\tau} \quad , \quad \Delta m \ll 1 
\end{eqnarray}
with $\tau=1$. The universal exponent $\tau =1$ for jump sizes $N^{-1/2} \ll \Delta m \ll 1$ results from superposed contributions from many overlaps, i.e.\ all scales, as illustrated in Fig.~\ref{fig:illustration}.

The asymptotics for large $\Delta m$ is controlled by small $q\ll 1$, i.e., by transitions between very distant states. Injecting the density of states $\nu(0)$ near $q=0$ and Taylor expanding in $q$ inside the exponential yields the estimate
\begin{eqnarray} 
 \rho(\Delta m)  &\approx&  \nu(0)\Delta m \int_0^\infty \rmd q \,\frac{\exp\Big(-\frac{(\Delta m)^2(1+q)}{4}\Big)}{\sqrt{4 \pi}} \nn\\
 &=&\frac{2 \nu(0)}{\sqrt{\pi}} \frac{e^{- (\Delta m)^2/4}}{(\Delta m)^{\tau'}}  \quad , \quad \Delta m \gg 1, \label{jumpdensity2}
\end{eqnarray}
with $\tau'=1$.
We see that avalanches with $\Delta m\gg 1$ ($\Delta M\gg \sqrt{N}$) are exponentially suppressed.

Plots at intermediate $\Delta m = O(1)$ are shown in Fig.~\ref{fig:illustration} for three different values of the external field. As no analytical closed form for $q(\hat x)$ is available, we have used approximations of the type
${\hat x}(q)=(a q+b q^2)/\sqrt{1-q}$ with $a=1.28$ and $b=-0.64$, proposed  in the literature~\cite{ParisiToulouse,FranzParisi}, and a sharp lower cutoff at~\cite{opperman} $q_{\rm min}(H)= 1.0 H^{2/3}$.

\subsection{Distribution of jumps: $H\neq 0$}

In the presence of a finite field $H$, Parisi's solution develops a plateau at low $\hat x$:
\begin{eqnarray}
q(\hat x<\hat x_m) = q_{m}(H), 
\end{eqnarray}
where $q_{m}(H)\approx 1.0 H^{2/3}$ and $\hat x_m\approx \nu(0) q_m(H)$ for small $H$, while $q(\hat x)$ is nearly unchanged for $\hat x>\hat x_m$.~\cite{MPV,ParisiToulouse, opperman} 
It is convenient to rewrite formula (\ref{rhoSK}) as
\be
\label{rhoSKwithH}
 \rho( \Delta m) = \theta(\Delta m) \Delta m  \int_{q_m(H)^-}^{q_c} \rmd q\,  \nu(q)  \frac{\exp\!\Big(-\frac{(\Delta m)^2}{4(q_c-q)}\Big)}{\sqrt{4\pi (q_c-q)}} . 
\ee
where the density of states $\nu(q)$ contains a piece $\delta(q-q_m) {x}_m/T$ when $q({x})$ exhibits a plateau at $x \leq x_m$, hence the notation $q_m^-$ in the integral.
Thus, the effect of a magnetic field is  to change the behavior of the jump distribution at large $\Delta m$, where it is now dominated by the plateau:
\begin{equation} 
\label{field_like_onestep}
\rho(\Delta m) =  \theta(\Delta m) \Delta m \, \hat x_m \frac{\exp\Big(-\frac{(\Delta m)^2}{4[1-q_m(H)]}\Big)}{\sqrt{4\pi [1-q_m(H)]}} .
\end{equation} 
Comparing with Eq.~(\ref{jumpdensity2}) we find an effective exponent $\tau'= -1$ (instead of 1) in the tail of the distribution. The formula (\ref{field_like_onestep}) holds only if we can neglect the contribution of the continuous part of $q(x)$. A simple comparison with the previous section shows that this holds when $\Delta m \gg \Delta m_H \sim 1/\hat x_m^{1/2}\sim  H^{-1/3}$. For $1\ll \Delta m\ll \Delta m_H$ the behavior crosses over to a formula similar to (\ref{jumpdensity2}) with $\tau'=1$.

Note that a small random field also produces a plateau in $q(x)$, and hence, we expect its effect on $\rho(\Delta m)$ to be rather similar to that of a uniform field.

\subsection{Interpretation for the SK model} 
\label{ss:InterpretationSK}
To find a natural interpretation of formula (\ref{rhoSK}) we consider what happens upon increasing $h$ from $h_1$ to $h_2$. If we take $h_{21}=h_2-h_1\ll 1$  we only need to consider the possibility that the ground state and the lowest-lying metastable state cross as we tune $h$, corrections  due to higher excited states being of order ${ O}(h_{12}^2)$.

We now argue that the disorder-averaged density of states of this two-level system is given by $\nu(q) \rmd q \, \rmd E$, where $\nu(q)$ was defined in Eq.~(\ref{nu_q}). Indeed, the definition of the overlap distribution $P(q)$ is
\begin{equation}
P(q)= \sum_{\alpha,\gamma} w_\alpha w_\gamma \delta(q-q_{\alpha\gamma})
\end{equation}
where $w_\alpha= \exp(-\beta F_\alpha)/\sum_\gamma \exp(-\beta F_\gamma)$ is the Gibbs weight of the metastable state $\alpha$. At low $T$ we can restrict to the two lowest states, which yield the leading-order term in 
\begin{equation}
P(q) = (1-T)\delta\big(q-q_c\big) + T \rho_1(q)+O(T^2)
\end{equation}
as 
\begin{eqnarray}
T\rho_1(q) &=& \int_0^\infty  \rmd E\, \nu(q,E) \frac{2 e^{-\beta E}}{(1+e^{-\beta E})^2}\nn\\
&=&T\nu(q,0) +O(T^{2}).
\end{eqnarray}
Here $\nu(q,E)$ is the joint probability density of overlap $q$ and free-energy difference $E$ between the ground and first excited state. Hence Eq.~(\ref{nu_q}) holds with $\nu(q)=\nu(q,0)$.

The two states differ in $N_{\rm fl}=N(1-q)/2$ flipped spins. In the SK model the magnetization is uncorrelated with the energy, and one thus expects the magnetization difference between the states to be a Gaussian variable of zero mean and variance (at fixed overlap) 
\bea
\langle \Delta m^2\rangle_q = 4 N_{\rm fl}/N= 2(1-q).
\eea
When $h$ increases the energy difference between the first excited state and the ground state changes from $E$ (for $h=h_1$) to $E-h_{21}\Delta m$, where $h_{21}:=h_{2}-h_{1}>0$.
Thus, if $\Delta m>0$, a jump at equilibrium occurs when $h_{21} = E/\Delta m$. For the shock probability per unit $h$ one thus expects 
\begin{eqnarray} 
\label{shocksrederived}
\rho(\Delta m) &=&\lim_{h_{21}\downarrow 0} \int_{q_m^-}^{q_c} \rmd q \int_0^\infty  \rmd E\, \nu(q,E)   \nn\\
&&\qquad \times\frac{\exp\! \Big(-\frac{(\Delta m)^2}{2\langle \Delta m^2\rangle_q}\Big)}{\sqrt{2\pi \langle\Delta m^2\rangle_q}}
\,\delta\!\left( h_{21}- \frac{E}{\Delta m}\right),\qquad 
\end{eqnarray}  
reproducing Eq.~(\ref{rhoSK}) upon integration over $E$. 

This argument strongly suggests that the joint density (per unit of $h$) of jumps with characteristics $q$ and $\Delta m$, is given by
\begin{equation} \label{rho_q}
\rho(\Delta m,q) = \theta(\Delta m) \Delta m\, \nu(q)  \frac{\exp\!\Big(-\frac{(\Delta m)^2}{2\langle \Delta m^2\rangle_q}\Big)}{\sqrt{2\pi \langle\Delta m^2\rangle_q}} .
\end{equation}
Integrating over $\Delta m$ we find the density of jumps with overlap $q$
as
\begin{equation}
\rho(q) = \sqrt{\frac{1-q}{\pi}} \,\nu(q),
\end{equation}
or for the density of flipped spins, $N_{\rm fl}=\frac{(1-q)N}{2}$ 
\begin{equation}
{\cal D}(N_{\rm fl})\rmd N_{\rm fl} = 
 \frac{1}{\sqrt{\pi}}\frac{2}{N} \sqrt{\frac{2 N_{\rm fl}}{N}}\,\nu\!\left(q=1-\frac{2N_{\rm fl}}{N}\right)  \rmd N_{\rm fl} .
\end{equation}
Let us now consider avalanches with $N_{\rm fl}\ll N$. Using Eq.~(\ref{qSK}), we find 
\begin{eqnarray}
\rho(q) = \frac{C}{\sqrt{\pi}}\frac{1}{1-q},
\end{eqnarray}
and the power law density 
\begin{eqnarray}
{\cal D}(N_{\rm fl}) =   \frac {C}{\sqrt{\pi}}\frac{1}{N_{\rm fl}^\rho},
\end{eqnarray}
with $\rho=1$.

\subsection{Comparison with numerical work}
For the SK model, there is no numerical study of equilibrium avalanches to date. However, in a pioneering work,  out-of-equilibrium avalanches at $T=0$ were studied numerically along the hysteresis loop~\cite{SKnumerics}, and found to 
exhibit criticality, i.e.\ a power-law distribution of magnetization jumps. The external field $H$ is increased adiabatically slowly until a single spin becomes unstable. The latter is flipped and triggers with finite probability  an avalanche of further spin flips, during which $H$ is kept fixed. The typical difference in applied magnetic field  between adjacent jumps scales as $N^{-1/2}$, which is the same scaling as in our calculation. During the avalanche a sequential single-spin-flip update was used to ensure the decrease of the total energy. Interestingly they observe the same scaling of the jumps of total magnetization, $\Delta M\sim N^{1/2}$, and the number of spin flips (which we assume to be of the same order as the number of spins that have flipped an odd number of times), $N_{\rm fl}\sim N$, as in our present calculation for equilibrium. It is interesting to note that this implies that a typical spin flips on the order of $N^{1/2}$ times along one branch of the hysteresis loop. 
A very similar density of avalanches with the same exponents $\tau=\rho=1$ and a crossover at $\Delta m\sim 1$,  as analytically obtained for the statics here,
was observed in the numerics. 
This similarity is  surprising since the states reached along the hysteresis curve are quite far from the ground state, as evidenced by the width of the hysteresis loop. Nevertheless, the visited states share an important feature with the ground state: self-organized criticality.
Indeed, the distribution of the local fields $h_i=\sum_{j\neq i} J_{ij} \sigma_j+H$, i.e., the energy cost to flip spin $i$ only, is observed to display a linear pseudogap~\cite{SKnumerics} as in the equilibrium~\cite{TAP}, marginally satisfying the minimal requirement for metastability. 

To understand better the relation between static and dynamic avalanches in the SK model, it would be useful to perform both equilibrium and dynamic simulations. In particular, it would be interesting to determine the prefactor of the power-law for the  density of jumps, which we have computed here for equilibrium, but which has not been determined in  Ref.~\onlinecite{SKnumerics}, because they normalized the jump density. It would also be interesting to compute the probability density of overlaps between states before and after an avalanche, and compare with the expression (\ref{rho_q}) derived in equilibrium. 

One could measure the joint density of overlaps and avalanche-sizes, 
\begin{equation}
\rho_{H}( \Delta m,q):= \left. \left< \delta\!\left(q-1+\frac{2 N_{\mathrm{fl}}}{N}\right) \delta\!\left(\Delta m-\frac{\Delta M}{\sqrt{N}}\right)\ \right> \right|_{H}\, ,
\end{equation}
where the average is taken for fixed external magnetic field (i.e., in practice for $H \in \left[ H-\delta H,H+\delta H\right]$, with $\delta H$ small). 
It would be interesting to check whether this joint density takes a form as in Eq.~(\ref{rho_q}) with $\langle\Delta m^2\rangle_q=2(1-q)$. In this case, this might allow to define a {\em dynamical} overlap-distribution $\nu(q)$ (to be interpreted as the $T=0$ limit of $P_{\rm dyn}(q)/T$). 

\section{Droplet argument in any $d$}
Let us now discuss the Edwards-Anderson model in dimension $d$. We first give a scaling argument to predict the avalanche exponent based on a droplet picture. Subsequently we will show how the previous result for the SK model can be recovered and interpreted in the same spirit.

To determine the first avalanche as the field is increased, we need information about the lowest-energy excitations of a given magnetization, which will scale inversely with the volume.
More precisely, we expect the lowest excitation energy for a droplet-like excitation of linear size $L$ to scale as 
\begin{eqnarray}
\label{dropletmin}
E_{\rm min}(L) \sim  \frac1{\nu_0}\frac{L^\theta}{V/L^{d_{\mathrm{f}}}}.
\end{eqnarray}
This is argued as follows: Standard droplet arguments~\cite{FisherHuse} stipulate that the lowest-energy excitation of linear size $L$, including a given spin, grows typically as $L^\theta$. These droplets are in general objects of fractal dimension ${d_{\mathrm{f}}}\leq d$. We thus assume that one can cover the system of volume $V$ by $V/L^{d_{\mathrm{f}}}$  droplets, and that they are uncorrelated. This implies the scaling (\ref{dropletmin}) for the droplet of minimal energy. The density $\rho_{0}$ of such single-droplet excitations near the ground state thus behaves as $\rho_{0} \mathrm dE = {\mathrm{d} E}/E_{\rm min}(L)$, or $\rho_{0} = \nu_{0}/ L^{\theta} \times V/L^{d_{\mathrm f}}$
 
The magnetization jump associated with the overturn of a droplet of size $L$ is assumed to scale as $L^{{d_{\mathrm{m}}}}$. Of course, $d_{\mathrm m}\leq d_{\mathrm f}$. The numerical study~\cite{Bouchaud} suggests that ${d_{\mathrm m}}$ is rather close to ${d_{\mathrm f}}$. 
We assume the total magnetization of droplets of size $L$  to be uncorrelated with the energy, and distributed as $P_L(\Delta M)=L^{-{d_{m}}}\psi_M(\Delta M/L^{{d_{m}}})$. 
In a vanishing field, low-energy droplets are believed to exist at all length scales.

We make the standard assumption that droplets at scale $L$ are uncorrelated from droplets at scales $\geq 2L$. By analogy with the reasoning given for the SK model, one argues that the density of avalanches per volume, per unit field $H$, and per unit magnetization change $\Delta M$ is given by
 \begin{eqnarray} 
\label{droplets}
\rho(\Delta M)  &\approx &\lim_{\delta H\downarrow 0} \frac{1}{V} \int_1^{\infty} \frac{{\mathrm{d} L}}{L} \int_0^\infty  \frac{{\mathrm{d} E}}{E_{\rm min}(L)} \\
&& \qquad \times \delta \bigg(\!\delta H-\frac{E}{\Delta M} \!\bigg) P_L(\Delta M).  \nn
\end{eqnarray}
Using the above expressions one  finds
 \begin{eqnarray} 
\rho(\Delta M)  
&\approx& 
 \frac{1}{(\Delta M)^\tau}\frac{\nu_0}{{d_{m}}}\int^{\infty}_0  \rmd z \,\psi_M(z) z^{\tau}, 
 \end{eqnarray} 
valid for $\Delta M\gg 1$, with the avalanche exponent 
\begin{eqnarray}
\tau = \frac{{d_{\mathrm{f}}}+\theta}{{d_{\mathrm{m}}}}.
\end{eqnarray}
This prediction is very general. As discussed in Ref.~\onlinecite{us}, it also gives reasonable predictions for elastic interfaces in random media. 
The formula was recently rediscovered in the context of the ferromagnetic phase of the random-field Ising model~\cite{GarelMonthus}, in which case ${d_{\mathrm{m}}}={d_{\mathrm{f}}}$, and thus $\tau =1+\theta/{d_{\mathrm{f}}}$.~\cite{footnoteGM}

It is interesting to point out the close analogy between the exact expression for the SK model (\ref{rhoSK}) and the  
 heuristic droplet argument (\ref{droplets}). In the SK model the role of spatial scale is played by the overlap distance $1-q$, and the logarithmic sum over scales $\int {\mathrm{d} L}/L$ goes over into an integral $\rmd q/(1-q)$. The equivalent of $E_{\rm min}(L)$ is given by the typical gap at distance $1-q$, which is known to be~\cite{FranzParisi} $\Delta_q = (1-q)^{1/2}$. Finally, the distribution of magnetizations at fixed droplet scale, $P_L(\Delta M)$ is given by
 $ P_L(\Delta M)=\frac{\textstyle\exp\big(-\frac{(\Delta m)^2}{4[1-q]}\big)}{\textstyle\sqrt{4\pi (1-q)}}$.
 Putting these elements together and substituting them into Eq.~(\ref{droplets}) without the volume normalization factor, one recovers expression (\ref{shocksrederived}) with $\nu(q)$ given in (\ref{qSK}). Note that changing variables from $H$ to $h= N^{1/2}H$ and $\Delta M$ to $\Delta m= N^{-1/2}\Delta M$ does not change the density of avalanche sizes per unit field and unit jump size.

In the presence of a finite field $H$, droplets are believed to be suppressed above a scale  $L_{H}\sim 1/H^{\gamma}$ (with $\gamma>0$). This implies that integration over droplet scales in Eq.~(\ref{droplets}) is cut off at $L_H$ leading to 
\begin{equation} 
\rho(\Delta M)  =
 \frac{1}{(\Delta M)^\tau}\frac{\nu_0}{{d_{\mathrm m}}}\int^{\infty}_{{\Delta M}/{L_H^{{d_{\mathrm{m}}}}}}  \rmd z \,\psi_M(z) z^{\tau}, 
 \end{equation} 
which cuts off the power-law decay of the avalanche-size distribution at $\Delta M\sim L_H^{{d_{\mathrm{m}}}}$.

At small but non-zero temperature we expect several effects. First, there is a thermal rounding of all the magnetization jumps, which is apparent in Eq.~(\ref{2ndcumulant}) and was discussed there. The equilibrium jumps are smeared out over an interval $\Delta h\sim T/\sqrt{T \chi_{\rm FC}}$. In order to be distinguishable from the sample-averaged increase of magnetization, the avalanches should be bigger than the latter $\Delta m\gg \Delta h \chi_{\rm FC}\sim \sqrt{T \chi_{\rm FC}}\sim T$. Above this scale, the avalanche distribution is unchanged for $T\ll T_c$.

\section{Conclusion} 

We have introduced a method based on replica techniques to compute the cumulants of the equilibrium magnetization in the SK model at different fields. From their non-analytic part we have extracted the distribution of magnetization jumps at $T=0$. It  exhibits an interesting power-law behavior, characteristic of the criticality of the spin-glass phase. We have also obtained a prediction of the avalanche-size exponent for spin glasses in any dimension using droplet arguments. We have compared with numerical simulations of the out-of-equilibrium dynamics of the SK model and found striking similarities with the static calculations presented here. 

It would be very interesting to investigate avalanches in small fields in realistic models, as the finite-range Edwards-Anderson model in 2 and 3 dimensions, to test some of the predictions that we obtained using droplet arguments. Furthermore, experimental measurements of power-law Barkhausen noise in spin glasses (e.g., by monitoring magnetization bursts~\cite{barkhausen,saclay}) could provide complementary insight to earlier investigations of equilibrium noise~\cite{noiseinspinglasses}. 

We expect similar critical response upon slow changes of system parameters in many other systems described by continuous replica-symmetry breaking, as, e.g., in various optimization problems (minimal vertex cover~\cite{vertexcover}, coloring~\cite{coloring}, and $k$-satisfiability~\cite{1stepstabilitykSAT}
 close to the satisfiability threshold, and in the UNSAT region at large $k$. 
Likewise, in models of complex economic systems, one expects a power-law distributed market response to changes in prices and stocks \cite{jpb}. 
Avalanches have also been predicted to occur in electron glasses with unscreened $1/r$ interactions, and have been studied numerically in detail in Ref.~\onlinecite{Palassini}. They find an avalanche exponent $\tau =3/2$, which is reminiscent of the value found for disordered interfaces and random-field systems at the upper critical dimension.
   
Finally, we comment on possible future avenues to explore. It would be interesting to study analytically the dynamics of avalanches in the SK model. In principle one could use methods developped for the aging dynamics~\cite{CugliandoloKurchan1993}. In the simplest framework, one studies relaxation from a random initial state, in which case the overlap between initial and final state  vanishes at large time. Presumably the hysteresis cycle selects a sequence of states which have non-trivial subsequent overlaps. This remains a challenge to describe analytically.
A more modest, but still non-trivial goal consists in describing the dynamics starting from an equilibrium state upon an increase of  magnetic field by a small amount $\sim N^{-1/2}$.

 It would be interesting to study whether the states visited dynamically along the hysteresis curve, and the avalanches triggered, have a relation with the marginal TAP states at high energies and their distinct soft modes\cite{MarginalStates}. It would also be interesting to analyze the multi-shock terms $O(|h|^{k>1})$ in the magnetization cumulants, allowing to determine whether there are correlations between successive jumps. 

We thank L.~Cugliandolo, S.~Franz, M.~Goethe, M.~Palassini, G.~Zarand, and G.~Zimanyi for interesting discussions. We thank KITP Santa Barbara for hospitality, while various parts of this work were accomplished. This research was supported  by ANR grant 09-BLAN-0097-01/2 and in part by the National Science Foundation under Grant No.\ NSF PHY05-51164.  

\appendix 
\begin{widetext}
\section{Zero'th order cumulant for the magnetization}\label{a:A}
Here we evaluate the contribution of $\phi^0$, Eq.~(\ref{c3}).
At $T=0$, one can set $H[\vec y] \to \max_i \{ y_i\}$, to simplify to 
\begin{eqnarray}
\overline{m_{ h_1} \ldots  m_{h_p}}^{J,c,(0)}
&=&   -(-T)^{p} 
\int \rmd^py\, \delta \Big(\sum_i \alpha_i y_i \Big) \partial_{h_1} \ldots  \partial_{h_p} \overline {\max \{\vec{y}+z \beta \vec h \sqrt{q(1)-q_m} \}}^z \qquad \nn\\
&=&  (-T)^{p-1} \sqrt{q(1)-q_m}
\int \rmd^py\, \delta \Big(\sum_i \alpha_i y_i \Big)  \partial_{h_2} \ldots  \partial_{h_p} \overline {z \prod_{i=2}^p\Theta\Big(y_1-y_i +\beta z \sqrt{q(1)-q_m}[h_1-h_i]\Big) }^z \qquad \nn\\
&=&   \sqrt{q(1)-q_m}^p
\int \rmd^py\, \delta \Big(\sum_i \alpha_i y_i \Big) \, \overline {z^p \prod_{i=2}^p\delta\Big(y_1-y_i +\beta z \sqrt{q(1)-q_m}[h_1-h_i]\Big) }^z \qquad \nn\\
&=&   \sqrt{q(1)-q_m}^p
\int \rmd y_1\, \overline { \delta \Big(y_1 +\sum_{i=2}^p \alpha_i \beta z\sqrt{q(1)-q_m}[h_1-h_i] \Big) \,  z^p  }^z \qquad \nn\\
&=&  \left[q(1)-q_m\right]^{p/2} \, \overline{z^p} = \left[2(q(1)-q_m)\right]^{p/2} \frac{ \left[(-1)^p+1\right] \Gamma \left(\frac{p+1}{2}\right)}{2\sqrt{\pi }}\ ,
\end{eqnarray}
which is the result given in the text.

\section{Magnetization cumulants to first order in the shock expansion} \label{s:app1}

Consider formula (\ref{startingpoint}). In the limit of $T\to 0$,  $\hat h= h/T$ becomes very large, and   we can approximate
$H(\vec{y})={\rm max}_i(y_i)$. The idea of the following calculation is that  taking a field derivative yields a derivative of $H(\vec{y})$, which is a $\delta$-function, eliminating one integration.

To evaluate  (\ref{startingpoint}), we start with the cross-term, and choose without loss of generality $\hat h_1 \le \hat h_2 \le \ldots \le  \hat h_p$:
\begin{equation}\label{A1}
  (-1)^{p}
  \partial_{\hat h_1} \ldots  \partial_{\hat h_p}  \frac{T}{2 }  \int_{q_m}^{q(u_c)} \rmd q\, \frac{\rmd\hat u(q)}{\rmd q}  \overline{ \prod_{i=1}^p \int_{-\infty}^\infty  \rmd y_i \, 
\delta\Big(\sum_i \alpha_iy_i\Big)    H\Big(\vec{y}+\vec{\hat h}A_M\Big)H\Big(\vec{y}+ \vec{\hat h}A_m\Big) }^{A_+,A-}\ ,
\end{equation}
where we have denoted $A_M:=\max(A_+,A_-)$ and $A_m:=\min(A_+,A_-)$. 
The derivatives can be written as 
\begin{equation}\label{A2}
 \partial_{\hat h_1} \ldots  \partial_{\hat h_p} \left[H\left(\vec{y}+\vec{\hat h}A_M\right)H\left(\vec{y}+ \vec{\hat h}A_m\right) \right]
 =\sum_{m=0}^p \sum_{\{j_{i}\},\{k_{i}\}}
 \partial_{\hat h_{j_1}} \ldots  \partial_{\hat h_{j_m}} H\left(\vec{y}+\vec{\hat h}A_M\right)\partial_{\hat h_{k_1}} \ldots  \partial_{\hat h_{k_{p-m}}} H\left(\vec{y}+ \vec{\hat h}A_m\right) \ , 
\end{equation}
where the sum is over partitions of the $p$ fields $\hat h_i$ into two groups of $m$ and $p-m$ fields with ${j_1}<\ldots < {j_m}$ and $k_1<\ldots <k_{p-m}$. The multiple derivative (with at least one derivative) of the first factor of $H$ can be written as 
\begin{eqnarray}\label{A3}
\partial_{\hat h_{j_1}} \ldots  \partial_{\hat h_{j_m}} H\left(\vec{y}+\vec{\hat h}A_M\right)
=(-1)^{m-1}A_M^m\prod_{\ell=2}^{m}\delta(y_{j_1}+\hat h_{j_1}A_M - y_{j_\ell}-\hat h_{j_\ell}A_M)  \prod_{i=1}^{p-m}\Theta(y_{j_1}+\hat h_{j_1}A_M - y_{k_i}-\hat h_{k_i}A_M), \nn\\
\end{eqnarray}
This equation is proven by noting that 
\begin{enumerate}
\item $\displaystyle \max(y_1, \ldots, y_p) = \sum_{i=1}^p  y_i \prod_{l\neq i} \theta(y_i-y_l)$. 
\item $\displaystyle \partial_{y_i} \max(y_1, \ldots, y_p) =  \prod_{l\neq i} \theta(y_i-y_l)$, since derivatives of the $\theta$-functions cancel in pairs. 
\item a further derivative of $\theta(y_i-y_l)$ w.r.t.\ $y_l$ gives $- \delta(y_i-y_l)$. 
\end{enumerate}
This result is a consequence of the fact that the maximum of $m$ variables depends on $p \leq m$ variables if and only if these are mutually equal. We note that this  expression is symmetric in the $\{ \hat h_{j_1},\ldots ,\hat h_{j_m}\}$, and that   a similar expression holds  for the second factor.

The terms $m=0$ and $m=p$ have to be considered separately, which we do now, starting with $m=p$: 
Using (\ref{A3}) and eliminating all the $\delta$-functions from the derivatives of $H$ yields \begin{eqnarray}
&& \prod_{i=1}^p \int_{-\infty}^\infty \rmd y_i \, 
\delta\left(\sum_i \alpha_iy_i\right)  \partial_{\hat h_1} \ldots  \partial_{\hat h_p} H\!\left(\vec{y}+\vec{\hat h}A_M\right) H\!\left(\vec{y}+ \vec{\hat h}A_m\right)\nn\\
&& =(-1)^{p-1} A_M^p \int_{-\infty}^\infty \rmd y_1 \,\delta\!\left(\sum_i \alpha_i [y_1+\hat h_1A_M-\hat h_iA_M]\right) {\rm max}_i \left\{y_1+\hat h_1A_M-\hat h_i(A_M-A_m)\right\}\nn\\
&& =(-1)^{p-1}A_M^p \left(A_M \sum_i \alpha_i \hat h_i - \hat h_1(A_M-A_m)\right), 
\end{eqnarray}
where to get to the last line we have used $\sum_i \alpha_i=1$ and $\min_i \{\hat h_i\}=\hat h_1$. 

Likewise the term $m=0$ gives
\begin{eqnarray}
&& \prod_{i=1}^p\int_{-\infty}^\infty \rmd y_i  \,
\delta\left(\sum_i \alpha_iy_i\right) H\!\left(\vec{y}+ \vec{\hat h}A_M\right) \partial_{\hat h_1} \ldots  \partial_{\hat h_p} H\!\left(\vec{y}+\vec{\hat h}A_m\right)\nn\\
&& =(-1)^{p-1}A_m^p \left(A_m \sum_i \alpha_i \hat h_i + \hat h_p(A_M-A_m)\right). 
\end{eqnarray}

Let us now discuss the terms $m=1, \ldots, p-1$. 
Consider 
\begin{eqnarray}
\lefteqn{\prod_{i=1}^p \int_{-\infty}^\infty \rmd y_i\,  \delta\Big(\sum_i \alpha_i y_i \Big)
 \partial_{\hat h_{j_1}} \ldots  \partial_{\hat h_{j_m}} H\left(\vec{y}+\vec{\hat h}A_M\right)\partial_{\hat h_{k_1}} \ldots  \partial_{\hat h_{k_{p-m}}} H\left(\vec{y}+ \vec{\hat h}A_m\right) }\nn\\
 &=& (-1)^{p-2}  A_M^{m}A_m^{p-m}\int_{-\infty}^\infty  \rmd y_{j_1}\int_{-\infty}^\infty \rmd y_{k_1}
  \prod_{i=1}^{p-m} \Theta\left( y_{j_1}+\hat h_{j_1}A_M - y_{k_1}-\hat h_{k_1}A_m - (A_M-A_m) \hat h_{k_i} \right) \nn\\
&& \hphantom{\int_{-\infty}^\infty(-1)^{p-2}  A_M^{m}A_m^{p-m}\int_{-\infty}^\infty  \rmd y_{j_1} \rmd y_{k_1}}
\times \prod_{l=1}^m \Theta\left( -\left[y_{j_1}+\hat h_{j_1}A_M - y_{k_1}-\hat h_{k_1}A_m\right] + (A_M-A_m)
 \hat h_{j_\ell} \right)\nn\\
&&\hphantom{\int_{-\infty}^\infty(-1)^{p-2}  A_M^{m}A_m^{p-m}\int_{-\infty}^\infty  \rmd y_{j_1} \rmd y_{k_1}}
\times\delta\left(\sum_\ell \alpha_\ell (y_{j_1}+A_M(\hat h_{j_1}-\hat h_{j_\ell}) +\sum_i \alpha_i (y_{k_1}+A_m(\hat h_{k_1}-\hat h_{k_i})\right) \nn\\
 &=& (-1)^{p-2}  A_M^{m}A_m^{p-m}\int_{-\infty}^\infty  \rmd y_{j_1}\int_{-\infty}^\infty \rmd y_{k_1}
 \Theta\left( y_{j_1}+\hat h_{j_1}A_M - y_{k_1}-\hat h_{k_1}A_m - (A_M-A_m){\rm max}_{i=1,\ldots ,p-m}
 \hat h_{k_i} \right) \nn\\
&& \hphantom{\int_{-\infty}^\infty(-1)^{p-2}  A_M^{m}A_m^{p-m}\int_{-\infty}^\infty  \rmd y_{j_1} \rmd y_{k_1}}
\times \Theta\left( -\left[y_{j_1}+\hat h_{j_1}A_M - y_{k_1}-\hat h_{k_1}A_m\right] + (A_M-A_m){\rm min}_{\ell=1,\ldots ,m}
 \hat h_{j_\ell} \right)\nn\\
&&\hphantom{\int_{-\infty}^\infty(-1)^{p-2}  A_M^{m}A_m^{p-m}\int_{-\infty}^\infty  \rmd y_{j_1} \rmd y_{k_1}}
\times\delta\left(\sum_\ell \alpha_\ell (y_{j_1}+A_M(\hat h_{j_1}-\hat h_{j_\ell}) +\sum_i \alpha_i (y_{k_1}+A_m(\hat h_{k_1}-\hat h_{k_i})\right) \label{A6}
\end{eqnarray}
Note that by going from the first to the second line, we have used the $\delta$-functions to fix $y_{j_l}=y_{j_1}+(\hat h_{j_1}-\hat h_{j_l}) A_M$, and $y_{k_i}=y_{k_1}+ (\hat h_{k_1}-\hat h_{k_i})A_m$. From the second to the third line we have used that $A_M-A_m \ge 0$ to simplify the 
products of $\Theta$-functions. 

The product of the two $\Theta$ functions implies that the contribution is non-zero only if the partitions satisfy $\hat h_{j_\ell}>\hat h_{k_i}$ for all $i,\ell$. 
Since we ordered $\hat h_1 \le \ldots \le \hat h_p$, this identifies the set of $\hat h_{k_i}$ to be $\{ \hat h_1, \ldots, \hat h_{p-m} \}$, and the set of $\hat h_{j_\ell}$ to be $\{ \hat h_{p-m+1}, \ldots, \hat h_{p}\}$. 

Making in (\ref{A6}) the shift of variables $y_{j_1}\to y_{j_1} + y_{k_1}$ eliminates $y_{k_1}$ from the $\Theta$ functions, and allows to do the integral over the latter, resulting into 
\begin{eqnarray}
(\ref{A6}) &=& (-1)^{p-2}  A_M^{m}A_m^{p-m}\int_{-\infty}^\infty  \rmd y_{j_1} \Theta\left( y_{j_1}+\hat h_{j_1}A_M -\hat h_{k_1}A_m - (A_M-A_m)\,{\rm max}_{i=1,\ldots ,p-m}
 \{ \hat h_{k_i} \} \right) \nn\\
&& 
\qquad \qquad \qquad \qquad \qquad \times \Theta\!\left( -\left[y_{j_1}+\hat h_{j_1}A_M -\hat h_{k_1}A_m\right] + (A_M-A_m)\, {\rm min}_{\ell=1,\ldots ,m}\{  \hat h_{j_\ell}\} \right) \nn\\
 &=& (-1)^{p-2}  A_M^{m}A_m^{p-m}\int_{-\infty}^\infty  \rmd y_{j_1} \Theta\!\left( y_{j_1} - (A_M-A_m)  \hat h_{p-m-1} \right) \Theta\!\left( -y_{j_1}+ (A_M-A_m) \hat h_{p-m} \right) \nn\\
 &=& (-1)^{p}  A_M^{m}A_m^{p-m} (A_M-A_m) \left(\hat h_{p-m+1}-\hat h_{p-m} \right) 
\end{eqnarray}
Putting all terms together, (\ref{A1}) becomes
\begin{eqnarray}
\lefteqn{ (-1)^p  \prod_{i=1}^p \int_{-\infty}^\infty \rmd y_i  \, 
\delta\Big (\sum_i \alpha_iy_i \Big)  \partial_{\hat h_1} \ldots  \partial_{\hat h_p} \left[H\left(\vec{y}+\vec{\hat h}A_M\right)H\left(\vec{y}+ \vec{\hat h}A_m\right)\right] }\nn\\
& =&-(A_M^{p+1}+A_m^{p+1}) \overline{h}   +(A_M-A_m)\left(- \hat h_p A_m^p+ \sum_{m=1}^{p-1} (\hat h_{p-m+1}-\hat h_{p-m}) A_m^{p-m}A_M^{m} +\hat h_1 A_M^p \right). 
\end{eqnarray}
where $\overline h:= \sum_i \alpha_i \hat h_i$. The first term $\sim \bar h$ disappears once we subtract the contributions from the non-crossed terms $\half \big[H(\vec{y}+\vec{\hat h}A_M)H(\vec{y}+ \vec{\hat h}A_M)\big]$ and $\half \big[H(\vec{y}+\vec{\hat h}A_m)H(\vec{y}+ \vec{\hat h}A_m)\big]$. This leads to the formula (\ref{overlined}) given in the text.

\section{Proof of Eq.~(\ref{Theorem})}
\label{app:proof}
Here we prove that for all sets of $\mu_a$ with replica indices $a=1,...,n$ the identity 
\begin{equation}
\label{Reltoprove}
 \sum_{i_a\in\{1,\dots,p\} |\sum_a\delta_{j,i_a}=n\alpha_j}^\prime \exp\left(\sum_{a=1}^n h_{i_a}\mu_a\right) =
 \frac{\int_{-\infty}^\infty \prod_{i=1}^p \rmd y_i \, \delta(\sum_{i=1}^p \alpha_iy_i) \prod_{a=1}^n\left[\sum_{i=1}^p\exp(h_{i}\mu_a+y_i)\right]}{\int_{-\infty}^\infty \prod_{i=1}^p\rmd y_i \, \delta\left(\sum_{i=1}^p \alpha_iy_i\right) \big[\sum_{i=1}^p\exp(y_i)\big]^{n}}.
\end{equation}
holds. By definition of the primed sum, the left hand side reduces to $1$ for $\mu_a=0$, in which case the idenitity is trivial. We now prove the identity by series expansion in $\mu_a\neq 0$.  
\end{widetext}
We define
\begin{equation}
K_{i}(\mu_a):=\frac{\exp\big(h_{i}\mu_a\big)}{\frac 1 p\sum_{j=1}^p\exp\big(h_{j}\mu_a\big)}-1,
\end{equation}
which has the property that $K_i(\mu_a=0)=0$, as well as $\sum_{i=1}^p K_i(\mu_a)=0$.
We can then write 
\begin{equation}
\exp(h_{i}\mu_a)=\left[1+K_{i}(\mu_a)\right]{\frac 1 p\sum_{j=1}^p\exp\big(h_{j}\mu_a\big)}.
\end{equation}
Analogously we define
\begin{eqnarray}
{\cal N}({\vec y}):=\frac{1}{p}\sum_{i=1}^p\exp(y_i),\\
\Delta_i({\vec y}):= \frac{\exp(y_i)}{{\cal N}({\vec y})}-1, 
\end{eqnarray}
so that  $\sum_{i=1}^p \Delta_{i}(y)=0$, and 
$\exp(y_i)={\cal N}({\vec y})\left[1+\Delta_i({\vec y})\right].$
With this one finds
\begin{eqnarray}
\lefteqn{\sum_{i=1}^p e^{y_i}e^{\mu_ah_i}}\\
&=&\sum_{i=1}^p {\cal N}({\vec y})\left[1+\Delta_i({\vec y})\right]\left[1+K_{i}(\mu_a)\right]{\frac 1 p\sum_{j=1}^p\exp\big(h_{j}\mu_a\big)}\nn\\
&=& {\cal N}({\vec y}) {\sum_{j=1}^p\exp\big(h_{j}\mu_a\big)} 
\left[1+\frac{1}{p}\sum_{i=1}^p \Delta_i({\vec y}) K_{i}(\mu_a)\right]\nn\\
&=&  {\cal N}({\vec y}) {\sum_{j=1}^p\exp\big(h_{j}\mu_a\big)} 
 \left[1+\sum_{i=1}^{p-1} \frac{\Delta_i({\vec y})-\Delta_p({\vec y})}{p} K_{i}(\mu_a)\right] . \nn
\end{eqnarray}
With this notation the identity (\ref{Reltoprove}) to be proven can be restated as
\begin{eqnarray}\nn
&&\sum_{i_a\in\{1,\dots,p\} |\sum_a\delta_{j,i_a}=n\alpha_j}^\prime\  \prod_{a=1}^n \left[1+K_{i_a}(\mu_a)\right] \\
&&\qquad\qquad=
 \frac{1}{N} \int\prod_{i=1}^p\rmd y_i \,\delta\left(\sum_{i=1}^p\alpha_i y_i\right) [{\cal N}({\vec y})]^{n} \nn\\
&&\qquad \qquad\qquad \times \prod_{a=1}^n \left[1+\sum_{i=1}^{p-1} K_{i}(\mu_a)\frac{\Delta_i({\vec y})-\Delta_p({\vec y}) }{p}\right],\qquad
\label{conjecture}
\end{eqnarray}
\nopagebreak
where we have divided  by the common factor $\prod_{a=1}^n \left(\frac1p \sum_{j=1}^{p}\exp(h_{j} \mu_a)\right)$
on both sides. The normalization $N$ is defined as 
\begin{equation}
N := \int\prod_{i=1}^p\rmd y_i \,\delta\left(\sum_{i=1}^p\alpha_i y_i\right) [{\cal N}({\vec y})]^{n}\ .
\end{equation}
This identity holds if and only if the coefficients of linearly independent products of factors of $K_i(\mu_a)$ are identical on both sides.  
Since $\sum^\prime$ is normalized,  the identity holds for $K_i(\mu_a)=0$, i.e.,~$\mu_a=0$.
We now consider products over factors $K_i$ with $i$ ranging over $1\leq i\leq p-1$, since $K_p= -\sum_{i=1}^{p-1}K_i$. Consider a product with $k_i$ factors $K_i(\mu_a)$ (with all $\mu_a$ different). The coefficient on the left-hand side is obtained from combinatoric considerations: A factor of $K_i$ either comes directly from a term $(1+K_i)$ in (\ref{conjecture}), or it results from a term $(1+K_p)$, upon replacing $K_p= -\sum_{i=1}^{p-1}K_i$. There are $\left(\begin{array}{c} k_i \\
r_i
\end{array}\right)
=k_i!/r_i!(k_i-r_i)!$ different ways to have   
$r_i$ factors of the latter origin (each contributing a factor $(-1)$ to the coefficient) and $k_i-r_i$ of the former. Then, $k_i$ of the $(n\alpha_i)$ $\mu$-indices with $i_a=i$ are already assigned, while the remaining $n\alpha_i -k_i$ indices $i$ need still to be assigned to a subset of the $n-\sum_{i=1}^{p-1} k_i$ replica with yet unfixed $i_a$. The number of possibilities to make disjoint assignments for all indices $i=1,...,p$ is
\begin{equation}
\label{Replicarelation}
\frac{(n-\sum_{i=1}^{p-1}k_i)!}{(n\alpha_p -\sum_{i=1}^{p-1}r_i)!\prod_{i=1}^{p-1}  
(n\alpha_i-k_i+r_i)!}.
\end{equation}
This is normalized by the number of assignments of $n\alpha_i$ indices $i$ to unconstrained replica $a$,
\pagebreak[3]
\begin{equation}
\frac{n!}{\prod_{i=1}^{p}  
(n\alpha_i)!}.
\end{equation}
Putting all elements together, the sought coefficient follows as 
\begin{eqnarray}
\label{Replicarelation-2}
C_{\{k_i\}}&\equiv& { \sum_{r_1=0 }^{k_1} \ldots \sum_{r_{p-1}=0 }^{k_{p-1}} } 
\frac{(n-\sum_{i=1}^{p-1}k_i)!}{(n\alpha_p -\sum_{i=1}^{p-1}r_i)!\prod_{i=1}^{p-1}  
(n\alpha_i-k_i+r_i)!}
\nn\\
&& \qquad\times\frac{\prod_{i=1}^{p}  
(n\alpha_i)!}{n!}  \prod_{i=1}^{p-1} (-1)^{r_i} \left(
\begin{array}{c} k_i \\
r_i
\end{array}
\right).
\end{eqnarray}\vspace{-8mm}
\begin{widetext}
On the other hand, the coefficient on the right-hand side is given by
\begin{eqnarray}
\label{C'}
C_{\{k_i\}}' &=&  \frac{1}{N} \int\prod_{i=1}^p\rmd y_i \,\delta\left(\sum_{i=1}^p\alpha_i y_i\right)
\prod_{i=1}^{p-1}\left[\frac{\Delta_i({\vec y})-\Delta_p({\vec y})}{p}\right]^{k_i}\, [{\cal N}(\vec y)]^n \nn\\ 
&=&\frac{\int_{-\infty}^\infty \prod_{i=1}^p \rmd y_i \, \delta(\sum_{i=1}^p\alpha_i y_i)
\prod_{i=1}^{p-1}  (e^{y_i}-e^{y_p})^{k_i}  \left( \sum_{i=1}^p e^{y_i}\right)^{n-\sum_{i=1}^{p-1}k_i}}
{\int_{-\infty}^\infty \prod_{i=1}^p \rmd y_i \, \delta(\sum_{i=1}^p\alpha_i y_i)  \left( \sum_{i=1}^p e^{y_i}\right)^n}.
\end{eqnarray}
Our task is to show that $C_{\{k_i\}} = C_{\{k_i\}}'$. We note that a priori $C_{\{k_i\}}$ is only defined for integer and positive $n\alpha_i$, while $C_{\{k_i\}}'$ is only defined for $n<0$, but not necessarily integer.
We will show that $C_{\{k_i\}}'$ has an analytic continuation to positive $n$ and $n\alpha_i$ which indeed coincides with $C_{\{k_i\}}$  where the latter is defined. Thus we interpret $C_{\{k_i\}}'$ as the analytical continuation of the replica expression, which can then be continued to $n\uparrow 0$.

Let us proceed by computing the numerator in Eq.~(\ref{C'}) (recalling that everywhere we assume $\sum_{i=1}^p\alpha_i=1$)
\begin{eqnarray}
B_{\{k_i\}} &:=&
\int_{-\infty}^\infty \prod_{i=1}^p \rmd y_i \, \delta\Big(\sum_{i=1}^p\alpha_i y_i\Big)
\prod_{i=1}^{p-1}  (e^{y_i}-e^{y_p})^{k_i}  \left( \sum_{i=1}^p e^{y_i}\right)^{n-\sum_{i=1}^{p-1}k_i} \nn\\
&=&
\int_{-\infty}^\infty \prod_{i=1}^{p-1} \rmd y_i'\, \rmd y_p \, \delta\Big(\sum_{i=1}^{p-1}\alpha_i y_i' +y_p\Big)
e^{n y_p}\prod_{i=1}^{p-1}  (e^{y_i'}-1)^{k_i}  \left( 1+\sum_{i=1}^{p-1} e^{y_i'}\right)^{n-\sum_{i=1}^{p-1}k_i}\nn\\
&=&
\int_{-\infty}^\infty \prod_{i=1}^{p-1} \rmd y_i' \, 
\prod_{i=1}^{p-1}  \left[e^{-n\alpha_i y_i'} (e^{y_i'}-1)^{k_i}\right]  \left( 1+\sum_{i=1}^{p-1} e^{y_i'}\right)^{n-\sum_{i=1}^{p-1}k_i}\nn\\
&=&
\frac{1}{\Gamma(-n+\sum_{i=1}^{p-1}k_i)} \int_0^\infty \frac{\rmd\lambda}{\lambda^{1+n-\sum_{i=1}^{p-1}k_i} }
 \int_{-\infty}^\infty \prod_{i=1}^{p-1} \rmd y_i' \, 
e^{-\lambda\left( 1+\sum_{i=1}^{p-1} e^{y_i'}\right)}\prod_{i=1}^{p-1}  \left[e^{-n\alpha_i y_i'} (e^{y_i'}-1)^{k_i}\right].  
\end{eqnarray}
Now we change variables to $a_i = e^{y_i'}$ and expand the powers,
\begin{eqnarray}
B_{\{k_i\}} &=&
\frac{1}{\Gamma(-n+\sum_{i=1}^{p-1}k_i)} \int_0^\infty \frac{\rmd\lambda\,  e^{-\lambda}}{\lambda^{1+n-\sum_{i=1}^{p-1}k_i} }
 \prod_{i=1}^{p-1} \int_{0}^\infty  \rmd a_i \sum_{r_i=0}^{k_i} 
\left(
 \begin{array}{c} k_i \\
r_i
\end{array}
\right)
(-1)^{r_i}a_i^{k_i-r_i-n\alpha_i -1} e^{-\lambda a_i} \nn\\
&=&{ 
\frac{1}{\Gamma(-n+\sum_{i=1}^{p-1}k_i)} \int_0^\infty \frac{\rmd\lambda \, e^{-\lambda}}{\lambda^{1+n-\sum_{i=1}^{p-1}k_i} }
 \prod_{i=1}^{p-1}  \sum_{r_i=0}^{k_i} \left(
 \begin{array}{c} k_i \\
r_i
\end{array}
\right) 
(-1)^{r_i} 
\frac{\Gamma(k_i-r_i-n\alpha_i)}{\lambda^{k_i-r_i-n\alpha_i } }  }\nn\\
&=& { \sum_{r_1=0 }^{k_1} \ldots \sum_{r_{p-1}=0 }^{k_{p-1}} }
\frac{\Gamma(-n\alpha_p +\sum_{i=1}^{p-1}r_i)}{\Gamma(-n+\sum_{i=1}^{p-1}k_i)} 
 \prod_{i=1}^{p-1}  \left(
 \begin{array}{c} k_i \\
r_i
\end{array}
\right) 
(-1)^{r_i} 
\Gamma(k_i-r_i-n\alpha_i)
\end{eqnarray}
Finally, we use the  relation $\Gamma(x) =\frac{\pi}{\sin(\pi x) \Gamma(1-x)}$ to rewrite  this (using that $k_i$ and $r_i$ are integers) as
\begin{equation}
\label{finalB}
B_{\{k_i\}} = \frac{(-1)^{p-1}\sin(n\pi)/\pi}{\prod_{i=1}^{p}\sin(n\alpha_i\pi)/\pi}  \sum_{\{0\leq r_i\leq k_i\}} 
 \frac{\Gamma(1+n-\sum_{i=1}^{p-1}k_i)}{\Gamma(1+n\alpha_p -\sum_{i=1}^{p-1}r_i)\prod_{i=1}^{p-1}  
\Gamma(1+n\alpha_i-k_i+r_i)} {  \prod_{i=1}^{p-1}  \left(
 \begin{array}{c} k_i \\
r_i
\end{array}
\right) (-1)^{ r_i} } \ .
\end{equation}
\end{widetext}
The ratio of $\Gamma$ functions in Eq.~(\ref{finalB}) can be continued to positive $n$. When $n$ and all $n \alpha_i$ become integers, the latter can be written as 
\begin{equation}
\frac{(n-\sum_{i=1}^{p-1}k_i)!}{{(n\alpha_p} -\sum_{i=1}^{p-1}r_i)!\prod_{i=1}^{p-1}  
{ (n\alpha_i}-k_i+r_i)!}. 
\end{equation}
Dividing by the normalization factor yields indeed $C_{\{k_i\}}$, which completes the proof.

Note that the normalization factor in the denominator of Eq.~(\ref{C'}), for $n\to 0$ is given  by
\begin{equation}
N(n\to 0) = \frac{(-1)^{p-1}\sin(n\pi)/\pi}{\prod_{i=1}^p\sin(n\alpha_i\pi)/\pi} \left[1+O(n)\right],
\end{equation}
which tends to $N \to 1/[(-n)^{p-1}\prod_i\alpha_i]$ when $n\uparrow 0$, as calculated previously  in Eq.~(\ref{normaliz}).


\end{document}